\documentclass[prd,showpacs,11pt]{revtex4}

\usepackage[dvips]{graphicx}
\usepackage{graphics}

\advance \topmargin by .4in

\begin{document}

\def\be{\begin{equation}}
\def\ee{\end{equation}}
\def\brr{\begin{eqnarray}}
\def\err{\end{eqnarray}}
\def\bmath{\begin{math}}
\def\emath{\end{math}}
\def\s{\sigma}
\def\p{\partial}
\def\no{\nonumber}
\def\vtheta{\vartheta}
\def\vphi{\varphi}

\today
\title{New Hamiltonian formalism and quasi-local conservation equations
of general relativity}

\author{Jong Hyuk Yoon\footnote{E-mail address :yoonjh@konkuk.ac.kr}}
\affiliation{
Department of Physics, Konkuk University, \\
Seoul 143-701, Korea \\
and \\
Enrico Fermi Institute, University of Chicago, \\
5640 S. Ellis Av., Chicago, IL 60637, U.S.A.}

\begin{abstract}

I describe the Einstein's gravitation of 3+1 dimensional spacetimes using the (2,2)
formalism without assuming isometries. In this formalism, quasi-local energy, linear
momentum, and angular momentum are identified from the four Einstein's equations of
the divergence-type, and are expressed geometrically in terms of the area of a
two-surface and a pair of null vector fields on that surface. The associated
quasi-local balance equations are spelled out, and the corresponding fluxes are
found to assume the canonical form of energy-momentum flux as in standard field
theories. The remaining non-divergence-type Einstein's equations turn out to be the
Hamilton's equations of motion, which are derivable from the {\it non-vanishing}
Hamiltonian by the variational principle. The Hamilton's equations are the evolution
equations along the out-going null geodesic whose {\it affine} parameter serves as
the time function. In the asymptotic region of asymptotically flat spacetimes, it is
shown that the quasi-local quantities reduce to the Bondi energy, linear momentum,
and angular momentum, and the corresponding fluxes become the Bondi fluxes. The
quasi-local angular momentum turns out to be zero for any two-surface in the flat
Minkowski spacetime. I also present a candidate for quasi-local {\it rotational}
energy which agrees with the Carter's constant in the asymptotic region of the Kerr
spacetime. Finally, a simple relation between energy-flux and angular momentum-flux
of a generic gravitational radiation is discussed, whose existence reflects the fact
that energy-flux always accompanies angular momentum-flux unless the flux is an
$s$-wave.

\end{abstract}


\maketitle

\renewcommand{\theequation}{\arabic{section}.\arabic{equation}}

\begin{section}{Introduction}{\label{intro}}
\setcounter{equation}{0}

It has been known for sometime that the Pleba{\~n}ski equation\cite{plebanski},
the self-dual Einstein's equation of 4 dimensional Euclidean space,
can be obtained as the large $n$ limit of the equations of motion of
a certain class of $sl(n)$-valued non-linear sigma models
in 2 dimensions\cite{qhan,qhan2,qhan3}. The equivalence of these equations
defined in two different dimensions is quite unexpected,
but if one realizes that large $n$ limit of the $sl(n)$ Lie algebra
is just the Lie algebra of area-preserving diffeomorphisms
of an auxiliary 2 dimensional surface\cite{bakas89,floratos89,fairlie90},
and that the equations of motion of
$sl(\infty)$-valued non-linear sigma models in 2 dimensions are in fact
partial differential equations on 4 dimensional space, then one might
be more comfortable with the idea of describing 4 dimensional
self-dual Einstein's gravity as a limit of a certain class of
2 dimensional field theories,
and can show that the two theories  are in fact identical.
This correspondence is supported further by the observation
that the Pleba{\~n}ski equation and the $sl(n)$-valued 2-dimensional non-linear
sigma models are both integrable.

One may be interested in extending this idea of describing 3+1 dimensional
theories from 1+1 dimensional perspective without the self-dual restriction
and in the Lorentzian regime.
There are several advantages of such a description, if it is possible at all.
They stem from the fact that 1+1 dimensional
field theories are usually more manageable than 3+1 dimensional ones,
both classically and quantum mechanically.
For example, a number of field theories in 1+1 dimensions are renormalizable,
due to the {\it dimensionlessness} of field variables in a naive power counting.
There would be an enormous gain if one ever succeeds in describing
the Einstein's gravitation in 3+1 dimensions as a limit of some kind of
1+1 dimensional field theories which eventually proves to be renormalizable.
This idea sounds strange but does not seem impossible,
since the renormalizability is highly sensitive to
the spacetime dimensions on which the theories are defined.

These reasonings led us to seek the possibility whether the Einstein's
gravity in 3+1 dimensions without the self-dual restriction
is describable as a 1+1 dimensional field
theory\cite{yoon92}.
The idea was simply to split a 3+1 dimensional spacetime
into a 1+1 dimensional base manifold and a 2 dimensional fibre space,
and write down the Einstein-Hilbert action.
Then the Einstein-Hilbert
action becomes 1+1 dimensional field theory action, where
the infinite dimensional group of diffeomorphisms
of 2 dimensional fibre space becomes the Yang-Mills gauge symmetry.
But this program was successful only in a formal sense, since the resulting
1+1 dimensional action did not seem much useful, which made the whole
idea of describing the Einstein's gravitation as 1+1 dimensional
field theory questionable.
The follow-up idea was to use the gauge freedom of the 3+1 dimensional
spacetime\cite{yoon93a,yoon99a,yoon99b,yoon99c,yoon01,yoon02}.
If one chooses one of the spacetime coordinates as the affine parameter of
the out-going null geodesic, then it turns out that
the 1+1 dimensional field theory description of the Einstein's gravitation
is simplified significantly.

The purpose of this paper is to present several unexpected results that I obtained
in the (2,2) fibre bundle description of the Einstein's gravitation, and discuss
their physical implications. First, I will present quasi-local balance equations of
energy, linear momentum, and angular momentum for an arbitrary compact two-surface,
which are just two-surface integrals of the four divergence-type equations
that are part of the Einstein's equations\cite{evans98,yoon01,yoon02}.
Quasi-local energy, linear momentum,
and angular momentum are expressed in the coordinate-independent and geometric way in
terms of the area of a two-surface and the in- and out-going null vector fields at
each point of that surface\cite{york93,york97}.
They are Bondi-like, since their
rates of changes are given by fluxes of the canonical form\cite{lan-lif2}
\be
T_{0\alpha}\eta^{\alpha} \sim \sum_{I} \pi_{I}{\pounds_{\eta}} q^{I},
\ee
where $\eta$ is an appropriate
vector field defined at each point of a two-surface.

Second, problems of defining quasi-local angular momentum and
associated rotational energy have been particularly subtle
issues\cite{prior77,geroch77,bramson78,ashtekar81,dray84,ashtekar91}.
This is due to the fact
that the very notion of rotation depends on the choice of the coordinates,
which implies that one can always remove the effects of rotation by working
in a co-rotating coordinate system. On the other hand, it is natural to
demand that the angular momentum and the rotational energy of
any compact two-surface in the flat Minkowski spacetime be zero.
It will be seen that our quasi-local angular momentum and rotational energy
not only become zero for any two-surface in the flat Minkowski spacetime,
but also reduce to the standard values of the total angular momentum and
the {\it Carter's constant} in the asymptotic region of the Kerr
spacetime, respectively\cite{carter68,hugh-pen72,carter79,felice80}.
In this sense, our quasi-local rotational energy may be regarded
as a quasi-local generalization of the Carter's constant of
a generic gravitational field.

Third, using the affine parameter of the out-going null geodesic as the
time coordinate, I will write down the Hamiltonian
of the Einstein's theory\cite{kuchar92}.
I will obtain the Hamilton's equations of motion from this Hamiltonian
using appropriate boundary conditions, which determine the time flows of the
field variables. Together with the quasi-local balance equations
(or the constraint equations depending on the signature
of the 3-dimensional hypersurface), it will be seen that the Hamilton's
equations of motion constitute the full Einstein's equations.

Finally, I will present a simple but general relation between
quasi-local energy-flux and angular momentum-flux of a generic gravitational
radiation that has no isometries.
It is a generalization of the well-known relation of
mass-loss and angular momentum-loss\cite{hartle70},
\be
{\delta U}={\omega \over m_{z}}{\delta L_{z} } \label{algae}
\ee
for small perturbations around a stationary and axi-symmetric spacetime,
where $\omega$ and  $m_{z}$ are the frequency and azimuthal angular momentum
of the perturbations, respectively. To my knowledge, such a relation
between these gravitational fluxes of the most general type has not
been discussed before, but it strongly indicates that
our identifications of fluxes are physically correct, since energy-flux
always carries angular momentum-flux unless the radiation is an $s$-wave.

This paper is organized as follows. In section \ref{kine},
I will introduce the kinematics of the (2,2) fibre bundle formalism,
and write down the Einstein's equations. Then I will discuss
1+1 dimensional gauge theory aspects of the Einstein's gravitation
of 3+1 dimensions from this fibre bundle point of view.

In section \ref{out}, I will study the four Einstein's equations
that are first-order in the derivatives along the out-going null
vector field.
These equations, which are the natural analogs of the Einstein's
constraint equations in the 3+1 formalism,
turn out to be divergence-type equations.
It is from the two-surface integrals of these equations
that one obtains quasi-local balance equations of gravitational energy,
linear momentum, and angular momentum.
I will also present quasi-local gravitational rotational energy.
The Carter's constant, which is usually interpreted
as a measure of intrinsic rotation of gravitational field,
is known to exist for a certain class of spacetimes that have two commuting
Killing symmetries, and for the Kerr spacetime, it is just the total angular
momentum squared.
Our quasi-local rotational energy reduces to the Carter's constant
for asymptotically Kerr spacetimes, as is shown in \ref{k2},
and therefore, may be regarded as a quasi-local generalization of
the Carter's constant to spacetimes that have no isometries.

In section \ref{hameq}, it will be shown that the remaining
Einstein's equations, which are second-order in the derivatives
along the out-going null vector field,
are the Hamilton's equations of motion derivable
from a non-vanishing Hamiltonian by the variational principle.
The details of this derivation are given in Appendix.
Thus, together with the quasi-local balance equations (or constraint equations
depending on the signature of the 3-dimensional hypersurface),
the Hamilton's equations of motion make up for the full Einstein's
equations in this formalism.

In section \ref{geom}, quasi-local energy, linear momentum, and angular momentum
of the previous sections will be expressed in the coordinate-independent and
geometric way, using the area of a two-surface and a pair of
null vector fields orthogonal to that surface.
Relative to a given background spacetime against which
these quasi-local quantities are measured, quasi-local energy
and linear momentum are given by the rates of changes of the area of
the two-surface along the in- and out-going null vector fields,
respectively, and quasi-local angular momentum associated
with a vector field $\xi$ is given
by two-surface integral of the projection of the {\it twist} of
the in- and out-going null vector fields onto $\xi$ modulo
a background-dependent subtraction term.

In section \ref{k2}, I will study the quasi-local balance equations at the null
infinity and show that they all agree with the well-known Bondi formulae
of energy-loss, momentum-loss, and angular momentum-loss.
In order to show these correspondences, it is necessary to find the asymptotic
fall-off rates of the metric and their derivatives near the null infinity,
using the affine parameter of the out-going null geodesic as the radial
coordinate\cite{yoon01,yoon02,wald84,penrose86,chrusciel02}.
I will present the asymptotic fall-off rates in this section.
It will be shown that the quasi-local rotational energy
in the asymptotic region of the asymptotically Kerr spacetimes
agrees with the Carter's constant of the Kerr spacetime.

In section  \ref{relation}, a general relation between
quasi-local energy-flux and angular momentum-flux will be
presented for a generic gravitational radiation.
When restricted to small perturbations around a stationary and axi-symmetric
spacetime, it will be shown that this relation reduces to the well-known
relation of mass-loss and angular momentum-loss in the perturbation theory
of the Kerr black hole\cite{hartle70}.

In the Appendix, I present in detail the derivation of the non-divergence type
Einstein's equations as the Hamilton's equations of motion associated with
a non-vanishing gravitational Hamiltonian.

\end{section}
\begin{section}{Kinematics}{\label{kine}}
\setcounter{equation}{0}

$\partial_{a}=\partial / \partial {y^{a}}$,
$\partial_{+}=\partial / \partial {u}$,
$\partial_{-}=\partial / \partial {v}$

In this section, I will introduce the kinematics of the (2,2)
fibre bundle formalism\cite{cho75,cho-freund75},
and write down the Einstein's equations.
This section serves mainly to fix the notations.
Let us consider the following line element
\begin{equation}
ds^2 = -2dudv - 2hdu^2 +\phi_{ab}
 \left( dy^a + A_{+}^{\ a}du +A_{-}^{\ a} dv \right)
\left( dy^b + A_{+}^{\ b}du +A_{-}^{\ b} dv
\right),    \label{yoon}
\end{equation}
where $+,-$ stands for $u,v$,
respectively
\cite{yoon93a,yoon99a,yoon99b,yoon99c,yoon01,
yoon02,din-sta78,din-sma80,new-unti62,new-tod80}.
In order to understand the geometry of this metric,
it is convenient to introduce the following vector fields
\begin{math}
\{ \hat{\partial}_{\pm} \}
\end{math}
defined as
\brr
& &
\hat{\partial}_{+}:=\partial_{+} - A_{+}^{\ a}\partial_{a},\label{pplus}\\
& &
\hat{\partial}_{-}:=\partial_{-} - A_{-}^{\ a}\partial_{a},\label{mminus}
\err
where
\be
\partial_{+}={\partial \over \partial u}, \hspace{.2in}
\partial_{-}={\partial \over \partial v}, \hspace{.2in}
\partial_{a}={\partial  \over \partial y^{a}}
\hspace{.2in}  (a=2,3).
\ee
The inner products of the vector fields $\{ \hat{\partial}_{\pm}, \partial_{a} \}$
are given by
\brr
& &
<\hat{\partial}_{+}, \ \hat{\partial}_{+}> = -2h, \hspace{.2in}
<\hat{\partial}_{+}, \ \hat{\partial}_{-}> = -1,  \hspace{.2in}
<\hat{\partial}_{-}, \ \hat{\partial}_{-}> =0, \nonumber\\
& &
<\hat{\partial}_{\pm}, \ \partial_{a}> =0, \hspace{.2in}
<\partial_{a}, \ \partial_{b}> =\phi_{ab}.
\err
The hypersurface $u={\rm constant}$ is an out-going null hypersurface
generated by $\hat{\partial}_{-}$ whose norm is zero.
The hypersurface $v={\rm constant}$ is generated by $\hat{\partial}_{+}$
whose norm is $-2h$, which can be either
negative, zero, or positive, depending on whether $\hat{\partial}_{+}$ is
timelike, null, or spacelike, respectively. The vector fields
\begin{math}
\{ \hat{\partial}_{\pm} \}
\end{math}
are called horizontal since they are orthogonal to $\{ \partial_{a}\}$,
and two dimensional section spanned by $\{ \hat{\partial}_{\pm} \}$
has the Lorentzian signature.
The intersection of two hypersurfaces $u,v={\rm constant}$ defines
a spacelike two-surface $N_{2}$ labeled by $\{ y^{a}\}$,
which is assumed to be compact with a positive-definite metric
$\phi_{ab}$ on it (see FIG. 1).
The metric $\phi_{ab}$ is decomposed into
the area element  ${\rm e}^{\sigma}$ and the conformal two-metric
$\rho_{ab}$ normalized to have a unit determinant
\be
\phi_{ab}={\rm e}^{\sigma}\rho_{ab}
\hspace{.2in}
( {\rm det}\ \rho_{ab}=1 ).                   \label{det}
\ee

\begin{figure}
\begin{center}
\vspace{-1in}
\includegraphics[width=5in]{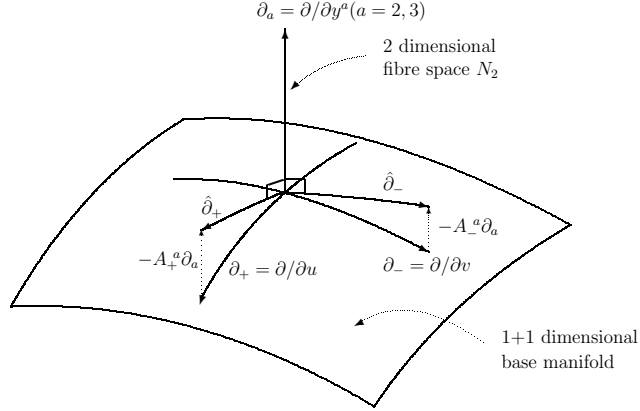}
\vspace{-3in}
\caption{\label{figure1} This figure shows the geometry of the (2,2) fibre
bundle splitting of 3+1 dimensional spacetime.
The 1+1 dimensional base manifold is
spanned by $\{ \partial_{\pm}\} $
and the 2 dimensional fibre space $N_{2}$ by $\{\partial_{a}\}$.
The horizontal vector fields $\{ \hat{\partial}_{\pm}\} $
are orthogonal to $N_{2}$, and $A_{\pm}^{\ a}$
are the connections valued in the Lie algebra of the diffeomorphisms of $N_{2}$. }
\end{center}
\end{figure}
For later uses, let us express the in-going null vector field $n$
and out-going null vector field $l$ in term of $\{ \hat{\partial}_{\pm} \}$.
They are given by
\brr
& & n= \hat{\partial}_{+} -  h  \hat{\partial}_{-}, \label{en}\\
& & l= \hat{\partial}_{-},                   \label{el}
\err
and satisfy the normalization condition
\be
< n, \ l > = -1.
\ee
Notice that $\partial / \partial v$ is either spacelike or null,
since its norm is given by
\be
<{\partial \over \partial v}, \ {\partial \over \partial v} >
={\rm e}^{\sigma}\rho_{ab}A_{-}^{\ a}A_{-}^{\ b} \geq 0.
\ee
The coordinate $v$ increases uniformly as $l$ evolves, since we have
\be
\pounds_{l}v=1.
\ee
In the gauge where $A_{-}^{\ a}=0$, $l$ is given by
\be
l={\partial \over \partial v},
\ee
which tells us that $v$ becomes the affine parameter of
the out-going null geodesic $l$.

The complete set of the vacuum Einstein's equations
are found to be\cite{yoon99b}
\begin{eqnarray}
(a) & &
{\rm e}^{\sigma} D_{+}D_{-}\sigma
+ {\rm e}^{\sigma} D_{-}D_{+}\sigma
+ 2{\rm e}^{\sigma} (D_{+}\sigma)(D_{-}\sigma)
- 2{\rm e}^{\sigma}(D_{-}h)(D_{-}\sigma)
- {1\over 2}{\rm e}^{ 2 \sigma}\rho_{a b}
F_{+-}^{\ \ a}F_{+-}^{\ \ b}                 \nonumber\\
& &
+{\rm e}^{\sigma} R_{2}
- h {\rm e}^{\sigma} \Big\{
(D_{-}\sigma)^{2}
-{1\over 2}\rho^{a b}\rho^{c d}
 (D_{-}\rho_{a c})(D_{-}\rho_{b d})\Big\}=0, \label{aa}\\
(b)  & &
-{\rm e}^{\sigma} D_{+}^{2}\sigma
- {1\over 2}{\rm e}^{\sigma}(D_{+}\sigma)^{2}
-{\rm e}^{\sigma}(D_{-}h) (D_{+}\sigma)
+{\rm e}^{\sigma}(D_{+}h)(D_{-}\sigma)
+2h {\rm e}^{\sigma}(D_{-}h)(D_{-}\sigma)  \nonumber\\
& &
+{\rm e}^{\sigma}F_{+-}^{\ \ a}\partial_{a}h
-{1\over 4}{\rm e}^{\sigma}\rho^{a b}\rho^{c d}
 (D_{+}\rho_{a c})(D_{+}\rho_{b d})
+\partial_{a}\Big( \rho^{a b}\partial_{b}h \Big) \nonumber\\
& &
+h\Big\{ - {\rm e}^{\sigma} (D_{+}\sigma)
  (D_{-}\sigma)
+{1\over 2}{\rm e}^{\sigma}\rho^{a b}\rho^{c d} (D_{+}\rho_{a c})
    (D_{-}\rho_{b d})
+{1\over 2}{\rm e}^{2\sigma}\rho_{a b}F_{+-}^{\ \ a}F_{+-}^{\ \ b}
-{\rm e}^{\sigma}R_{2} \Big\} \nonumber\\
& &
+h^{2}{\rm e}^{\sigma}\Big\{
(D_{-}\sigma)^{2}
-{1\over 2}\rho^{a b}\rho^{c d}
(D_{-}\rho_{a c}) (D_{-}\rho_{b d})\Big\}=0,  \label{bb}\\
(c) & &
2{\rm e}^{\sigma}(D_{-}^{2}\sigma) +
{\rm e}^{\sigma} (D_{-}\sigma)^{2}
    + {1\over 2}{\rm e}^{\sigma}\rho^{a b}\rho^{c d} (D_{-}\rho_{a c})
    (D_{-}\rho_{b d})=0,                 \label{cc}\\
(d) & &
D_{-}\Big( {\rm e}^{2\sigma}
\rho_{a b}F_{+-}^{\ \ b}\Big)
- {\rm e}^{\sigma}\partial_{a}(D_{-}\sigma)
- {1\over 2}{\rm e}^{\sigma}\rho^{b c}\rho^{d e}
    (D_{-}\rho_{b d})(\partial_{a}\rho_{c e})
+ \partial_{b} \Big(
{\rm e}^{ \sigma}\rho^{b c}D_{-}\rho_{a c} \Big) \nonumber\\
& & =0,\label{dd}\\
(e) & &
-D_{+}\Big( {\rm e}^{2\sigma}
\rho_{a b}F_{+-}^{\ \ b}\Big)
-{\rm e}^{\sigma}\partial_{a} (D_{+}\sigma )
   -{1\over 2}{\rm e}^{\sigma}\rho^{b c}\rho^{d e}
   (D_{+}\rho_{b d})(\partial_{a}\rho_{c e})
+\partial_{b}\Big(
{\rm e}^{ \sigma}\rho^{b c}D_{+}\rho_{a c} \Big) \nonumber\\
& &
+2h{\rm e}^{\sigma}\partial_{a}(D_{-}\sigma)
+h{\rm e}^{\sigma}\rho^{b c}\rho^{d e}
   (D_{-}\rho_{b d})(\partial_{a}\rho_{c e})
+2{\rm e}^{\sigma}\partial_{a}(D_{-}h)
-2\partial_{b}\Big(h
{\rm e}^{ \sigma}\rho^{b c}D_{-}\rho_{a c} \Big) \nonumber\\
& &
=0,                              \label{ee}\\
(f) & &
-2 {\rm e}^{ \sigma}D_{-}^{2} h
-2{\rm e}^{ \sigma} (D_{-}h)(D_{-}\sigma)
+ {\rm e}^{ \sigma}D_{+}D_{-}\sigma
+ {\rm e}^{ \sigma}D_{-}D_{+}\sigma
+ {\rm e}^{ \sigma} (D_{+}\sigma)(D_{-}\sigma) \nonumber\\
& &
+ {1\over 2}{\rm e}^{ \sigma}\rho^{a b}\rho^{c d}
  (D_{+}\rho_{a c})(D_{-}\rho_{b d})
+ {\rm e}^{2 \sigma}\rho_{a b}
    F_{+-}^{\ \ a}F_{+-}^{\ \ b}
 -2h{\rm e}^{ \sigma} \Big\{
   D_{-}^{2} \sigma +{1\over 2}(D_{-}\sigma)^{2} \nonumber\\
& &
+{1\over 4}\rho^{a b}\rho^{c d}
   (D_{-}\rho_{a c})(D_{-}\rho_{b d})\Big\}=0,  \label{fff}\\
(g)   & &
h\Big\{ {\rm e}^{\sigma} D_{-}^{2} \rho_{ab}
- {\rm e}^{\sigma}\rho^{c d}(D_{-}\rho_{a c})(D_{-}\rho_{b d})
+{\rm e}^{\sigma}(D_{-}\rho_{a b})(D_{-}\sigma) \Big\}  \nonumber\\
& &
-{1\over 2}{\rm e}^{\sigma} \Big(
D_{+}D_{-}\rho_{a b} + D_{-}D_{+}\rho_{a b} \Big)
+{1\over 2}{\rm e}^{\sigma} \rho^{c d}\Big\{
(D_{-}\rho_{a c})(D_{+}\rho_{b d})
+(D_{-}\rho_{b c})(D_{+}\rho_{a d}) \Big\}  \nonumber\\
& &
-{1\over 2}{\rm e}^{\sigma}\Big\{
(D_{-}\rho_{a b})(D_{+}\sigma)
+(D_{+}\rho_{a b})(D_{-}\sigma)  \Big\}  \nonumber\\
& &
 +{\rm e}^{\sigma}(D_{-}\rho_{a b})(D_{-}h)
+{1\over 2}{\rm e}^{2 \sigma}\rho_{a c}\rho_{b d}
 F_{+-}^{\ \ c}F_{+-}^{\ \ d}
-{1\over 4}{\rm e}^{2 \sigma}\rho_{a b}
 \rho_{c d}F_{+-}^{\ \ c}F_{+-}^{\ \ d}=0. \label{ggg}
\end{eqnarray}
Here $R_{2}$ is the scalar curvature of $N_{2}$, and
the diff$N_{2}$-covariant derivatives are given by\cite{yoon92,yoon99a},
\begin{eqnarray}
& &F_{+-}^{\ \ a}=\partial_{+} A_{-} ^ { \ a}-\partial_{-}
  A_{+} ^ { \ a} - [A_{+}, \ A_{-}]_{\rm L}^{a},  \label{field}\\
& &D_{\pm}\sigma = \partial_{\pm}\sigma
-[A_{\pm}, \ \sigma]_{\rm L},         \label{get}\\
& &D_{\pm}h= \partial_{\pm}h - [A_{\pm}, \ h]_{\rm L},   \label{eichid}\\
& &D_{\pm}\rho_{a b}=\partial_{\pm}\rho_{a b}
   - [A_{\pm}, \ \rho]_{{\rm L}a b}.    \label{rhod}
\end{eqnarray}
In general, the diff$N_{2}$-covariant derivative of a tensor density
$f_{a b\cdots }$ with weight $w$ with respect to the diffeomorphisms
of $N_{2}$ is given by
\be
D_{\pm}f_{a b\cdots}= \partial_{\pm}f_{a b\cdots}
-[A_{\pm}, \ f]_{{\rm L}a b\cdots},               \label{covdiff}
\ee
where the bracket $[A_{\pm}, \ f]_{{\rm L} a b\cdots }$ is the Lie
derivative of $f_{ab \cdots}$ along $A_{\pm}:=A_{\pm}^{\ a}\partial_{a}$,
\be
[A_{\pm}, \ f]_{{\rm L}a b\cdots}
:=A_{\pm}^{\ c}\partial_{c}f_{ab\cdots}
+f_{cb\cdots}\partial_{a}A_{\pm}^{\ c}
+f_{ac\cdots}\partial_{b}A_{\pm}^{\ c}
+\cdots
+w (\partial_{c}A_{\pm}^{\ c})f_{ab\cdots}.           \label{liebra}
\ee
For instance, the diff$N_{2}$-covariant derivatives of
the area element ${\rm e}^{\sigma}$ and the conformal metric
$\rho_{a b}$ which are scalar and tensor density with weight $1$ and $-1$
with respect to the diff$N_{2}$ transformations, respectively,
are given by
\brr
& & D_{\pm}{\rm e}^{\sigma}=\partial_{\pm}{\rm e}^{\sigma}
-A_{\pm}^{\ c}\partial_{c}{\rm e}^{\sigma}
-(\partial_{a} A_{\pm}^{\ a}) {\rm e}^{\sigma},      \label{walleye}\\
& & D_{\pm}\rho_{a b}=\partial_{\pm} \rho_{a b}
-A_{\pm}^{\ c}\partial_{c} \rho_{a b}
-\rho_{c b}\partial_{a}A_{\pm}^{\ c}
-\rho_{a c}\partial_{b}A_{\pm}^{\ c}
+ (\partial_{c}A_{\pm}^{\ c})\rho_{a b}.               \label{bass}
\err
If one uses the Leibniz rule in (\ref{walleye}), then one has
\brr
D_{\pm}{\sigma}
& = & \partial_{\pm}{\sigma} -A_{\pm}^{\ c}\partial_{c}{\sigma}
-\partial_{a} A_{\pm}^{\ a}          \nonumber\\
& = & \partial_{\pm}\sigma -[A_{\pm}, \ \sigma]_{\rm L}, \label{crappie}
\err
which is just the equation (\ref{get}).

The spacetime integral of the {\it scalar} curvature of the metric
(\ref{yoon}) is given by
\begin{equation}
I= \int \! \! du \, dv \, d^{2}y \, L
+ {\rm surface}  \ {\rm integrals},       \label{bareact}
\end{equation}
where $L$ is
given by\cite{yoon93a,yoon99a,yoon99b,yoon99c,yoon01,yoon02}
\begin{eqnarray}
& & L = -{1\over 2}{\rm e}^{2 \sigma}\rho_{a b}
  F_{+-}^{\ \ a}F_{+-}^{\ \ b}
  +{\rm e}^{\sigma} (D_{+}\sigma) (D_{-}\sigma)
  -{1\over 2}{\rm e}^{\sigma}\rho^{a b}\rho^{c d}
 (D_{+}\rho_{a c})(D_{-}\rho_{b d})
 -{\rm e}^{\sigma} R_2            \nonumber\\
& & -2{\rm e}^{\sigma}(D_{-}h)(D_{-}\sigma)
- h {\rm e}^{\sigma}(D_{-}\sigma)^2
+{1\over 2}h  {\rm e}^{\sigma}\rho^{a b}\rho^{c d}
 (D_{-}\rho_{a c})(D_{-}\rho_{b d}).    \label{barelag}
\end{eqnarray}
Each term in (\ref{barelag}) is
manifestly invariant under the diffeomorphisms of $N_{2}$, since
the $\{y^{a}\}$-dependence of each term is completely hidden
in the diff$N_{2}$-covariant derivatives.
In this sense one may regard $N_{2}$ as
a kind of {\it internal} space as in Yang-Mills theory,
with the infinite dimensional group of diffeomorphisms of $N_{2}$ as
the associated gauge symmetry.
Thus, the Einstein's gravitation of 3+1 dimensional spacetimes
is describable as 1+1 dimensional Yang-Mills type gauge theory
interacting with 1+1 dimensional scalar fields $\sigma$, $h$,
and non-linear sigma fields $\rho_{ab}$ whose interactions are
dictated by the above Lagrangian density $L$.
If one uses the diff$N_{2}$ gauge freedom so that $A_{-}^{\ a}=0$,
then the metric (\ref{yoon}) becomes identical
to the metric of the null hypersurface formalism studied
in \cite{new-unti62}.
In this paper, however, I shall retain the $A_{-}^{\ a}$ field,
since its presence will make the coordinate choice less restrictive
and the diff$N_{2}$-invariant Yang-Mills type gauge theory aspect
more transparent.
\end{section}

\begin{section}{A Set of Quasi-local Balance  Equations}{\label{out}}
\setcounter{equation}{0}

Notice that the equations (\ref{aa}), (\ref{bb}) and (\ref{ee})
are partial differential equations that are {\it first-order}
in $D_{-}$ derivatives. Therefore it is of particular interest to study
these equations, since they are the analogs of
the Einstein's constraint equations in the standard 3+1 formalism.
Thus, in this (2,2) formalism,
the {\it natural} vector field that defines the evolution
is $D_{-}$.
Then the momentum variables
\begin{math}
\pi_{I}=\{ \pi_{h},\pi_{\sigma}, \pi_{a}, \pi^{a b} \}
\end{math} \
conjugate to the configuration variables
\begin{math}
q^{I}=\{ h, \sigma,   A_{+} ^ { \ a}, \rho_{a b} \}
\end{math} \
are defined as
\begin{equation}
\pi_{I}:={\partial L\over \partial (D_{-}{q}^{I}) }.
\label{momenta}
\end{equation}
They are found to be
\begin{eqnarray}
& &\pi_{h}=-2 {\rm e}^{\sigma}(D_{-}\sigma),   \label{pih}\\
& &\pi_{\sigma} = -2 {\rm e}^{\sigma} (D_{-}h)
       -2h {\rm e}^{\sigma} (D_{-}\sigma)
     + {\rm e}^{\sigma} (D_{+}\sigma),   \label{pisigma}  \\
& &\pi_{a}={\rm e}^{2 \sigma} \rho_{a b}F_{+-}^{\ \ b}, \label{pia} \\
& &\pi^{a b}=
h{\rm e}^{\sigma} \rho^{a c}\rho^{b d}(D_{-}\rho_{c d})
-{1\over 2}{\rm e}^{\sigma} \rho^{a c}\rho^{b d}
(D_{+}\rho_{c d}).                            \label{definition}
\end{eqnarray}
Conversely, one can express $D_{-}$ derivatives of the configuration variables
in terms of the conjugate momenta as follows,
\brr
& & D_{-}h =  -{1\over 2}{\rm e}^{-\sigma}\pi_{\sigma}
 +{1\over 2}D_{+}\sigma
 +{1\over 2}h{\rm e}^{-\sigma}\pi_{h},  \label{dmh}\\
& & D_{-}\sigma
= -{1\over 2}{\rm e}^{-\sigma}\pi_{h}, \label{dmsig}\\
& & F_{+-}^{\ \ a}
={\rm e}^{-2\sigma}\rho^{a b}\pi_{b},    \label{dmf}\\
& & D_{-}\rho_{a b}={1\over h}{\rm e}^{-\sigma}
\rho_{a c}\rho_{b d}\pi^{c d}
+{1\over 2h}D_{+}\rho_{a b}.             \label{dmrho}
\err
Notice that  $\pi^{a b}$ is traceless
\begin{equation}
\rho_{ab}\pi^{a b}=0,
\end{equation}
due to the identities
\begin{equation}
\rho^{ab}D_{\pm}\rho_{ab}=0                   \label{module}
\end{equation}
which are direct consequences of the condition
\be
{\rm det}\ \rho_{ab}=1.        \label{hammerhead}
\ee
The Hamiltonian density $H_{0}$ is given by\cite{yoon01,yoon02}
\begin{eqnarray}
H_{0}&:=& \pi_{I}D_{-}{q}^{I} - L  \nonumber\\
&=& H + {\rm total}\ {\rm  divergences}, \label{ok}
\end{eqnarray}
where $H$ is
\begin{eqnarray}
& & H  =  -{1\over 2}{\rm e}^{-\sigma}\pi_{\sigma}\pi_{h}
+ {1\over 4}h{\rm e}^{-\sigma}\pi_{h}^{2}
-{1\over 2}{\rm e}^{-2\sigma}\rho^{a b}\pi_{a}\pi_{b}
+{1\over 2h}{\rm e}^{-\sigma}
\rho_{a c}\rho_{b d}\pi^{a b}\pi^{c d}    \nonumber\\
& &
+{1\over 2}\pi_{h}(D_{+}\sigma)
+{1\over 2h}\pi^{a b}(D_{+}\rho_{a b})
+{1\over 8h}{\rm e}^{\sigma}\rho^{a b} \rho^{c d}
(D_{+}\rho_{a c}) (D_{+}\rho_{b d})
+{\rm e}^{\sigma}R_{2}.        \label{tilde}
\end{eqnarray}
Notice that $H$ and $H_{0}$ are Hamiltonian densities that differ
by total divergences only.
In terms of the canonical variables,
the first-order equations (\ref{aa}), (\ref{bb}), and
(\ref{ee}) can be written as, after a little algebra,
\begin{eqnarray}
({\rm i}) \hspace{.5cm} & &
\pi^{a b}D_{+}\rho_{a b}   + \pi_{\sigma}D_{+}\sigma
-h D_{+} \pi_{h}
-\partial_{+}\Big(
     h\, \pi_{h} + 2 {\rm e}^{\sigma}D_{+}\sigma \Big) \nonumber\\
& &
+\partial_{a}\Big(
    h\, \pi_{h}A_{+}^{ \ a}
+ 2A_{+}^{ \ a}{\rm e}^{\sigma}D_{+}\sigma
 + 2h {\rm e}^{-\sigma}\rho^{a b}\pi_{b}
 +2\rho^{a b}\partial_{b}h \Big) =0,         \label{qenergy}\\
({\rm ii}) \hspace{.5cm} & &
H - \partial_{+}\pi_{h}
+ \partial_{a} \Big(
A_{+}^{\ a}\pi_{h}
+ {\rm e}^{-\sigma}\rho^{a b} \pi_{b} \Big)=0,  \label{qmomentum}\\
({\rm iii}) \hspace{.5cm} & &
\partial_{+}\pi_{a}
-\partial_{b}(A_{+}^{\ b}\pi_{a})
-\pi_{b}\partial_{a}A_{+}^{\ b}
-\pi_{\sigma}\partial_{a}\sigma
+ \partial_{a}\pi_{\sigma} - \pi_{h}\partial_{a}h
     - \pi^{b c}\partial_{a}\rho_{b c}   \nonumber\\
& &      + \partial_{b}( \pi^{b c}\rho_{a c})
     + \partial_{c}( \pi^{b c}\rho_{a b})
     - \partial_{a}( \pi^{b c}\rho_{b c})=0.  \label{qangular}
\end{eqnarray}
Notice that  (\ref{qenergy}) and (\ref{qmomentum})
are divergence-type equations of the following form\cite{evans98}
\be
A+ \partial_{+}B+ \partial_{a}C^{a}= 0.        \label{rainbow}
\ee
One can also express (\ref{qangular}) as another divergence-type equation.
If we contract (\ref{qangular})
by an arbitrary function $\xi^{a}$ of $\{ v, y^{b} \}$ such that
\begin{equation}
\partial_{+} \xi^{a}=0,                 \label{xicon}
\end{equation}
then we have
\begin{equation}
\pi^{a b}{\pounds_{\xi}} \rho_{a b}
+\pi_{\sigma}{\pounds_{\xi}} \sigma
+\pi_{h}{\pounds_{\xi}} h
+\pi_{a}{\pounds_{\xi}} A_{+}^{\ a}
-\partial_{+}( \xi^{a}\pi_{a} )
+\partial_{a}\Big(
-\xi^{a}  \pi_{\sigma} + 2 \pi^{a b} \xi^{c} \rho_{b c}
+A_{+}^{\ a} \xi^{b} \pi_{b} \Big)
=0,                                  \label{qangular2}
\end{equation}
which is in the same form as (\ref{rainbow}).
Here ${\pounds_{\xi}}f_{a b\cdots }$ is the Lie derivative
defined in (\ref{liebra}),
\be
\pounds_{\xi}f_{a b\cdots}=[\xi, \ f]_{{\rm L}a b\cdots}
\hspace{.2in}
(\xi:=\xi^{a}\partial_{a}).
\ee
Integrals of the divergence-type equations (\ref{qenergy}),
(\ref{qmomentum}), and (\ref{qangular2}) over a compact two-surface $N_{2}$
become, after normalizing by $1/ 16\pi$,
\begin{eqnarray}
& & {\partial \over \partial u} U(u,v)
={1\over 16\pi}  \oint \! d^{2}y \, \Big(
\pi^{a b}D_{+}\rho_{a b} +\pi_{\sigma}D_{+}\sigma
-h D_{+} \pi_{h}
   \Big),              \label{enflux} \\
& &  {\partial \over \partial u} P(u,v)
={1 \over 16\pi} \oint \! d^{2}y \, H,  \label{momflux}  \\
& & {\partial \over \partial u} L(u,v;\xi)
={1 \over 16\pi}\oint \! d^{2}y \, \Big(
\pi^{a b}{\pounds}_{\xi} \rho_{a b}
+\pi_{\sigma}{\pounds}_{\xi} \sigma
-h {\pounds}_{\xi} \pi_{h}
-A_{+}^{\ a}{\pounds}_{\xi}\pi_{a} \Big),   \label{angflux}
\end{eqnarray}
where we used the identities
\brr
& &
\oint \! d^{2}y \,  \pi_{h} {\pounds_{\xi}}h
= - \oint \! d^{2}y \,  h {\pounds_{\xi}} \pi_{h},   \label{shark}\\
& &
\oint \! d^{2}y \,  \pi_{a} {\pounds_{\xi}}A_{+}^{\ a}
= - \oint \! d^{2}y \,  A_{+}^{\ a}
{\pounds_{\xi}} \pi_{a}.                       \label{dolphin}
\err
Here $U(u,v)$, $P(u,v)$, and $L(u,v;\xi)$ are two-surface
integrals defined as
\begin{eqnarray}
& & U(u,v):= {1 \over 16\pi}\oint d^{2}y \,   \Big(
h\, \pi_{h} + 2 {\rm e}^{\sigma} D_{+}\sigma \Big)
+ \bar{U},               \label{enint}\\
& &
P(u,v):={1 \over 16\pi} \oint \! d^{2}y \, ( \pi_{h} )
+ \bar{P},                \label{momint}\\
& &
L(u,v;\xi):={1 \over 16\pi} \oint \! d^{2}y \,
(\xi^{a}\pi_{a}) + \bar{L}
\hspace{.2in} (\partial_{+} \xi^{a}=0),   \label{angint}
\end{eqnarray}
where $\bar{U}$, $\bar{P}$, and $\bar{L}$ are undetermined subtraction
terms that satisfy the conditions
\begin{equation}
{\partial \bar{U}\over \partial u}
={\partial \bar{P}\over \partial u}
={\partial \bar{L}\over \partial u} =0.    \label{subtract}
\end{equation}
Notice that choices of subtraction terms are not unique,
since it is the subtraction terms that {\it define} the references against
which these quasi-local quantities are measured.
A natural criterion for the ``right'' choice of subtraction
terms would be that values of quasi-local quantities reproduce
``standard'' values in the well-known limiting cases, but otherwise
they can be chosen arbitrarily.

\begin{figure}
\vspace{-1in}

\begin{center}
\[       
\hspace{-.8in}
\includegraphics[width=5in]{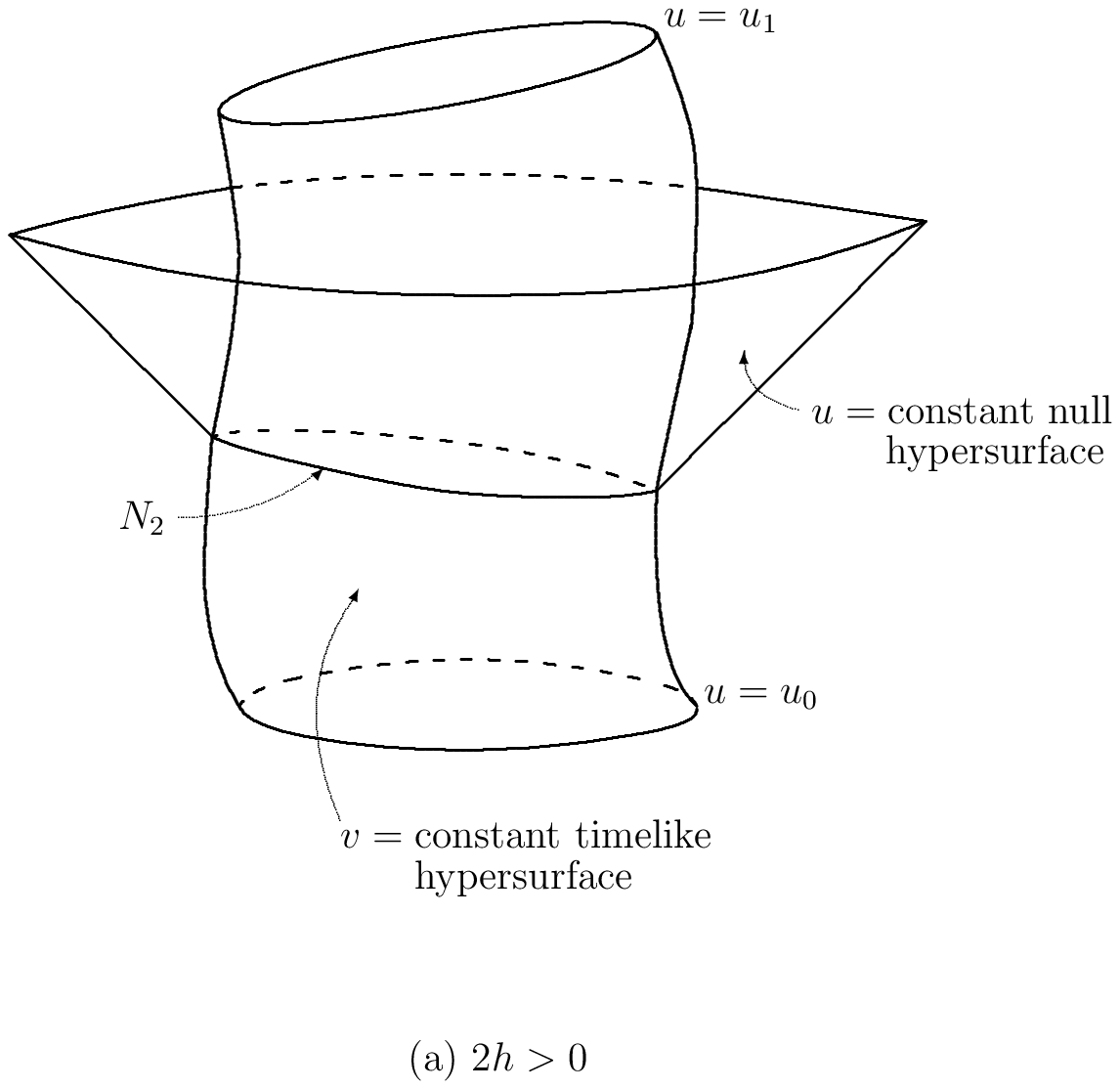}
\hspace{-2in}
\includegraphics[width=5in]{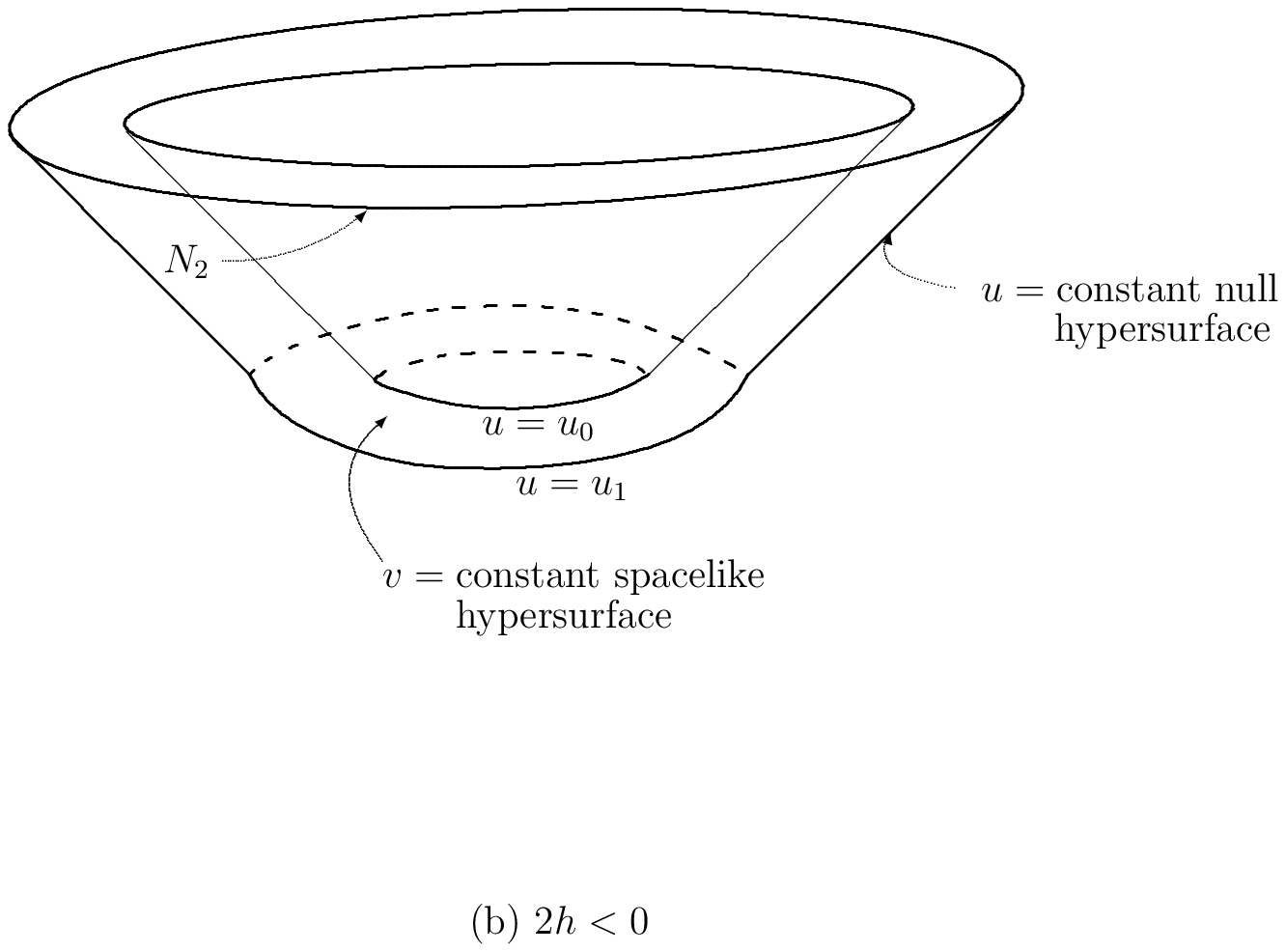}
\]
\vspace{-3in} \caption{\label{figure2}
(a) This figure shows the spacetime geometry in the region where
$2h>0$, and $v$=constant hypersurface is timelike.
(b) This figure shows the spacetime geometry in
the region where $2h<0$, and $v$=constant hypersurface is spacelike.
In both cases $N_{2}$ is represented by a circle.}

\end{center}
\end{figure}

Let us notice that the integrand of the r.h.s. of (\ref{angflux})
assumes the canonical form of angular momentum-flux,
\be
T_{0a}\xi^{a} \sim \sum_{I}\pi_{I}{\pounds}_{\xi}q^{I}, \label{tangential}
\ee
where $\xi$ is tangent to the two-surface $N_{2}$.
One can also put the r.h.s. of (\ref{enflux}) into the canonical form
of energy-flux,
\be
T_{0\alpha }\eta^{\alpha}
\sim \sum_{I}\pi_{I}\partial_{+}q^{I},        \label{also}
\ee
where $\eta^{\alpha}=\delta_{+}^{\ \alpha}$.
To show this, let us contract (\ref{qangular})
with $A_{+}^{\ a}$ and integrate over $N_{2}$. Then we have
\begin{equation}
\oint \! d^{2}y \, \Big(
A_{+}^{\ a} \partial_{+}\pi_{a} \Big)
=\oint \! d^{2}y \, \Big(
\pi^{a b}{\pounds}_{A_{+}} \rho_{a b}
+\pi_{\sigma}{\pounds}_{A_{+}} \sigma
-h {\pounds}_{A_{+}} \pi_{h} \Big).       \label{xxx}
\end{equation}
If we use the equation (\ref{xxx}) and the diff$N_{2}$-covariant
derivative $D_{+}$ defined in (\ref{covdiff}), then (\ref{enflux})
can be written as
\begin{equation}
{\partial \over \partial u} U(u,v)
={1\over 16\pi}  \oint \! d^{2}y \, \Big(
\pi^{a b}  \partial_{+} \rho_{a b}
+\pi_{\sigma}\partial_{+}\sigma
- h \partial_{+} \pi_{h}
-A_{+}^{\ a} \partial_{+}\pi_{a} \Big),   \label{enflux1}
\end{equation}
where the r.h.s. indeed assumes the canonical form of energy-flux\cite{lan-lif2}.

In the region where
\begin{math}
\hat{\partial}_{+}
\end{math} \
is timelike ($2h>0$), the equations (\ref{momflux}), (\ref{angflux}),
and (\ref{enflux1}) are quasi-local balance equations that relate
the instantaneous rates of changes of two-surface integrals at a given $u$-time
to the associated net fluxes across the timelike tube generated by
$\hat{\partial}_{+}$ (see FIG. 2a). Let us remark that,
unlike the Tamburino-Winicour's quasi-local conservation
equations that are ``weak'' conservation equations
since  the Ricci flat conditions (i.e. the full vacuum Einstein's
equations) were assumed in their derivation\cite{wini66},
our quasi-local balance equations are ``strong''
conservation equations in that only four Einstein's equations
of the divergence-type were used in the derivation.

In the region where
\begin{math}
\hat{\partial}_{+}
\end{math} \
is spacelike ($2h<0$), the vector field $\partial / \partial u$ is spacelike,
\be
<{\partial \over \partial u}, \ {\partial \over \partial u}> = -2h + {\rm
e}^{\sigma}\rho_{a b}A_{+}^{\ a}A_{+}^{\ b} > 0,
\ee
so that $u$ is the radial coordinate (see FIG. 2b).
Then the equations (\ref{qenergy}), (\ref{qmomentum}), and
(\ref{qangular2}) are constraint equations rather than  balance equations,
and each equation splits into a divergence term and a source term.
The source term either assumes the
canonical form of energy-momentum ``density'' $\Sigma
\pi_{I}\pounds_{\xi}{q}^{I}$ and $\Sigma \pi_{I}\partial_{+}{q}^{I}$ as in Eqs.
(\ref{tangential}) and (\ref{also}), or is given by the Hamiltonian ``density'' $H$
in (\ref{tilde}). Notice that the source term is not ``flux'' but ``density'',
since the $v={\rm constant}$ hypersurface is now a 3-dimensional spacelike
hypersurface. These constraint equations describe how the quasi-local quantities at
a given $u$-radius change as the radius $u$ changes on a given spacelike hypersurface.
Thus, the difference of two-surface integrals evaluated on two
successive two-spheres on a given spacelike hypersurface is given by the
3-dimensional spatial integral of the ``density'' over the region between the
two-spheres of interest. This is exactly what happens in Maxwell's theory, where the
Gauss law constraint is given by
\be \vec{\nabla} \cdot \vec{E} = 4\pi \rho.
\label{gauss}
\ee
If one integrates (\ref{gauss}) over a 3-dimensional spacelike
region $V$ whose boundaries are two-spheres $S_{1}$ and $S_{2}$, then
one has
\be
\oint_{S_{1}} \! \! E_{n} \, d a -\oint_{S_{2}} \! \!
E_{n} \, d a =  4\pi \int_{V} \! \rho \, \, d v.           \label{intgauss}
\ee
Thus, in Maxwell's theory, the difference of two-surface integrals
of the ``momentum'' $E_{n} $
on two successive two-spheres is given by the spatial integral
of the charge ``density'' over the volume between the two-spheres. In this paper,
however, we shall be concerned with the case $2h>0$ only, and the case $2h\leq 0$
will be discussed elsewhere\cite{isola}.

If we introduce a function $W_{R}(u,v)$ defined as
\be
W_{R}(u,v)
:={1\over 16\pi}\int^{u}_{0}\!\! \! du \! \oint \! d^{2}y \,
A_{+}^{\ a}\partial_{+}\pi_{a},   \label{w1}
\ee
then we have
\be
{\partial \over \partial u}W_{R}(u,v)
={1\over 16\pi}\oint \! d^{2}y \,
A_{+}^{\ a}\partial_{+}\pi_{a},          \label{w2}
\ee
so that (\ref{enflux1}) can be written as
\begin{equation}
{\partial \over \partial u} \Big\{ U(u,v) + W_{R}(u,v) \Big\}
={1\over 16\pi}  \oint \! d^{2}y \, \Big(
\pi^{a b}  \partial_{+} \rho_{a b}
+\pi_{\sigma}\partial_{+}\sigma  - h \partial_{+} \pi_{h}
\Big).                          \label{w3}
\end{equation}
Notice that the r.h.s. of (\ref{w2}) has the form
\be
{\partial \over \partial u}W_{R}(u,v)
\sim \sum_{a}\Omega^{a}\partial_{+} L_{a},
\ee
where
\brr
& & \Omega^{a} \sim A_{+}^{\ a} ,         \label{ural}\\
& & L_{a} \sim {1\over 16\pi}\pi_{a},     \label{ironshark}
\err
which represents the work done per unit $u$-time by changing
the angular momentum $L_{a}$ of the system that has the angular
velocity  $\Omega^{a}$. From this perspective, the equation (\ref{w2})
is the work done on $N_{2}$
per unit $u$-time by changing the angular momentum density $\pi_{a}/16\pi$
of gravitational field that has the angular velocity  $A_{+}^{\ a}$
at each point of $N_{2}$. This observation suggests that $W_{R}(u,v)$ be
identified as the quasi-local {\it rotational} energy of $N_{2}$.
Indeed, as is shown in section \ref{k2},
$W_{R}(u,v)$ reduces to the Carter's constant for the asymptotically
Kerr spacetimes, the total angular momentum
squared\cite{carter68,hugh-pen72,carter79,felice80}.
This is a strong indication that supports
our identification of $W_{R}(u,v)$ as the rotational energy of
gravitational field in the circumstances where {\it no} isometries
are present.

In the limit where $A_{+}^{\ a}$ is independent of $u$ such that
\be
\partial_{+}A_{+}^{\ a}=0,            \label{w4}
\ee
$W_{R}(u,v)$ becomes
\be
W_{R}(u,v)
={1\over 16\pi}\oint_{u=u} \! d^{2}y \, (A_{+}^{\ a}\pi_{a})
-{1\over 16\pi}\oint_{u=0} \! d^{2}y \, (A_{+}^{\ a}\pi_{a}).
              \label{w5}
\ee

\end{section}

\begin{section}{Hamilton's equations of motion}{\label{hameq}}
\setcounter{equation}{0}

Let us define the Hamiltonian $K$ as the integral of
$H$ in (\ref{tilde}),
\be
K:= \int \! \! du
\oint\!\! d^{2}y \, \Big\{
H+\lambda \, ({\rm det}\ \rho_{a b} - 1) \Big\},  \label{fundam}
\ee
where $\lambda$ is a Lagrange multiplier that enforces
the unimodular condition (\ref{hammerhead}). In the Appendix,
we have shown that the equations (\ref{cc}), (\ref{dd}), (\ref{fff}),
and (\ref{ggg}) are the Hamilton's equations of motion
\brr
& & D_{-}q^{I}=
  {\delta K \over \delta \pi_{I}}, \label{defm}\\
& & D_{-}\pi_{I}=
  -{\delta K \over \delta q^{I}},   \label{defe}
\err
where $\{ \pi_{I}$, $q^{I}\} $ are
\be
\pi_{I}=\{ \pi_{h}, \pi_{\sigma}, \pi_{a}, \pi^{a b} \},
\hspace{.2in}
q^{I}=\{ h, \sigma,   A_{+} ^ { \ a}, \rho_{a b} \},
\ee
assuming the boundary conditions
\be
\delta \sigma =\delta \rho_{a b} = 0       \label{bcon}
\ee
at the endpoints of the $u$-integration.
Thus, together with the divergence-type Einstein's equations
(\ref{aa}), (\ref{bb}), and (\ref{ee}), from which follow the integral equations
(\ref{enflux}), (\ref{momflux}), and (\ref{angflux}) that may be interpreted
as either the quasi-local balance equations or constraint equations depending on the
signature of the 3-dimensional hypersuface, the Hamilton's equations of motion
(\ref{defm}) and (\ref{defe}) make up for the complete set of
the vacuum Einstein's equations.
Thus, in the (2,2) fibre bundle formalism, the Einstein's equations split
into twelve first-order Hamilton's equations of motion dictating the
evolution along the out-going null geodesic and
the four quasi-local balance equations or the constraint equations
that implement the Hamilton's evolution equations.

\end{section}

\begin{section}{Geometrical interpretations}{\label{geom}}
\setcounter{equation}{0}

Two-surface integrals (\ref{enint}), (\ref{momint}), and (\ref{angint})
can be given geometric interpretations in terms of
the area of $N_{2}$ and null vector fields orthogonal to $N_{2}$.
In order to see this, it is necessary to recall the definitions of
the in- and out-going null vector
fields $n$ and $l$ given by (\ref{en}) and (\ref{el}), respectively.

\begin{subsection}{Quasi-local energy}{\label{geom1}}

Let us observe that, apart from the subtraction term $\bar{U}$,
(\ref{enint}) can be written as
the Lie derivative of the area $A$ of $N_{2}$ along $n$. Notice that
we have
\brr
\oint d^{2}y \, \Big( h\, \pi_{h} + 2 {\rm e}^{\sigma} D_{+}\sigma \Big)
& = &  2 \oint d^{2}y \,
{\rm e}^{\sigma} \Big (D_{+}\sigma - h D_{-}\sigma \Big) \nonumber\\
& = &  2 \oint d^{2}y \, {\pounds}_{n}{\rm e}^{\sigma}.  \label{triv}
\err
But one has
\be
\oint d^{2}y \, {\pounds}_{n}{\rm e}^\sigma
= {\pounds}_{n} {\cal A},                          \label{inter}
\ee
where ${\cal A}$ is given by
\be
{\cal A}=\oint d^{2}y \,{\rm e}^\sigma.
\ee
The identity (\ref{inter}) follows from the fact that
the order of $d^{2}y$ integration and the Lie derivation ${\pounds}_{n}$
is interchangeable, since the in-going null vector field
$n$ is orthogonal to $N_{2}$. Thus we have
\be
U(u,v)={1 \over 8\pi} {\pounds}_{n}{\cal A} +\bar{U}.
\label{bailfish}
\ee
In order to fix $\bar{U}$, it is necessary to introduce a reference spacetime.
In principle, the reference spacetime can be chosen arbitrarily,
provided that the pull-back of the background metric
to $N_{2}$ is the same as ${\rm e}^{\sigma}\rho_{ab}$.
If we denote the coordinates of the reference spacetime as
$(\bar{u}, \bar{v}, y^{a})$, then its metric can be written as
\begin{equation}
d\bar{s}^2
= -2d\bar{u}d\bar{v} - 2\bar{h}d\bar{u}^2 +{\rm e}^{\sigma} \rho_{ab}
\left( dy^a + \bar{A}_{+}^{\ a}d\bar{u} +\bar{A}_{-}^{\ a} d\bar{v} \right)
\left( dy^b + \bar{A}_{+}^{\ b}d\bar{u} +\bar{A}_{-}^{\ b} d\bar{v}
\right),                         \label{yoonref}
\end{equation}
where
\begin{math}
\{ \bar{h}, \bar{A}_{\pm}^{\ a} \}
\end{math}
are the embedding degrees of freedom of $N_{2}$ into the reference spacetime.
The vector fields $\{ \bar{n}$, $\bar{l} \}$
\brr
& &
\bar{n}=\Big( {\partial \over \partial \bar{u}}
- \bar{A}_{+}^{\ a}{\partial \over \partial y^{a}} \Big)
-\bar{h}  \Big(
{\partial \over \partial \bar{v}}
- \bar{A}_{-}^{\ a}{\partial \over \partial y^{a}}\Big), \label{enbar}\\
& &
\bar{l}=\Big( {\partial \over \partial \bar{v}}
- \bar{A}_{-}^{\ a}{\partial \over \partial y^{a}} \Big)
                             \label{elbar}
\err
are null with respect to the background metric, and satisfy
the same normalization conditions as before,
\be
<\bar{n}, \ \bar{n} >_{\rm ref}=0, \hspace{.2in}
<\bar{l}, \ \bar{l} >_{\rm ref}=0, \hspace{.2in}
<\bar{n}, \ \bar{l} >_{\rm ref}=-1.        \label{grouse}
\ee
If $\bar{U}$ is chosen as
\be
\bar{U}:=-{1 \over 8\pi} {\pounds}_{\bar{n}}{\cal A} \label{ubar}
\ee
such that it satisfies the $u$-independent condition
\be
{\partial \bar{U} \over \partial u}=0,    \label{grouper}
\ee
then (\ref{bailfish}) becomes
\be
U(u,v)
= {1 \over 8\pi} {\pounds}_{(n-\bar{n})}{\cal A}.  \label{eninta}
\ee
This expression is entirely geometrical, stating that $U(u,v)$
is determined by the rate of change of the area ${\cal A}$ of $N_{2}$
along the difference $n-\bar{n}$ of the in-going null geodesics,
and becomes zero when
\be
n=\bar{n}.
\ee

\end{subsection}

\begin{subsection}{Quasi-local linear momentum}{\label{geom2}}

One can also express $P(u,v)$ geometrically. Let us notice that
\be
{1 \over 16\pi} \oint \! d^{2}y \, ( \pi_{h} )
=-{1 \over 8\pi} \oint \! d^{2}y \, {\rm e}^{\sigma} D_{-}\sigma
=-{1 \over 8\pi} \oint \! d^{2}y \, {\rm e}^{\sigma}
 {\pounds}_{l} \sigma
=-{1 \over 8\pi}{\pounds}_{l}{\cal A}.     \label{mominta}
\ee
Therefore, if we choose the subtraction term  $\bar{P}$
as
\be
\bar{P}:={1 \over 8\pi}{\pounds}_{\bar{l}}{\cal A} \label{pbar}
\ee
such that it satisfies the $u$-independent condition
\be
{\partial \over \partial u} \bar{P}=0, \label{pbarcon}
\ee
then (\ref{momint}) becomes
\be
P(u,v)=-{1 \over 8\pi}{\pounds}_{(l-\bar{l})}{\cal A}.  \label{momintb}
\ee
Thus, $P(u,v)$ is given by the minus of the rate of change of
the area ${\cal A}$ of $N_{2}$ along the difference $l-\bar{l}$
of the out-going null geodesics, and becomes zero when
\be
l=\bar{l}.
\ee

\end{subsection}

\begin{subsection}{Quasi-local angular momentum}{\label{geom3}}

Let us now find the geometrical expression of $L(u,v;\xi)$.
If we notice that the Lie bracket of $n$ and $l$
is given by
\be
[n,\ l ]_{\rm L}
=-F_{+-}^{ \ \ a}\partial_{a} + (D_{-}h) l,                  \label{brown}
\ee
then we have
\brr
{1 \over 16\pi} \oint \! d^{2}y \,
(\xi^{a}\pi_{a})
&=&{1 \over 16\pi} \oint \! d^{2}y \,
{\rm e}^{ \sigma} \xi_{a} F_{+-}^{\ \ a}   \nonumber\\
&=&-{1 \over 16\pi} \oint \! d^{2}y \,
{\rm e}^{ \sigma} \xi_{a}  [n, \ l ]_{\rm L}^{a}  \nonumber\\
&=&-{1 \over 16\pi} \oint \! d^{2}y \,
{\rm e}^{ \sigma} <\xi,  \ [n, \ l ]_{\rm L}>,  \label{twist}
\err
where we used
\be
\xi_{a}= {\rm e}^{ \sigma} \rho_{a b} \xi^{b},     \label{xidef}
\ee
and the fact that $l$ is orthogonal to $\xi$,
\be
l^{a} \xi_{a}=0.                             \label{marlin}
\ee
If we choose the subtraction term $\bar{L}$ as
\be
\bar{L}
:={1 \over 16\pi} \oint \! d^{2}y \, {\rm e}^{ \sigma}
< \xi, \ [\bar{n}, \ \bar{l} ]_{\rm L}>_{\rm ref},  \label{elref}
\ee
and require that the condition
\be
{\partial  \over \partial u} \bar{L} =0  \label{elbarr}
\ee
hold, then (\ref{angint}) becomes
\be
L(u,v;\xi)
=-{1 \over 16\pi} \oint \! d^{2}y \,
{\rm e}^{ \sigma} \Big( < \xi, \ [n, \ l ]_{\rm L}>
- <\xi, \ [\bar{n}, \ \bar{l} ]_{\rm L}>_{\rm ref} \Big). \label{deff}
\ee
Thus, $L(u,v;\xi)$ is given by the integral over $N_{2}$ of the projection
of the twist $[n,\ l]_{\rm L}$ onto $\xi=\xi^{a}\partial_{a}$ modulo
the subtraction term $\bar{L}$.
Notice that $\xi^{a}$ is an arbitrary function of $\{v, y^{b}\}$, so that
the vector field $\xi$ need not satisfy any Killing's equations.
Thus, $\xi$ is an arbitrary vector field tangent to $N_{2}$,
defining the direction of rotation at each point of that surface.

If  $L(u,v;\xi)$ is to be regarded as
an acceptable candidate of quasi-local angular momentum,
its value must be zero for any two-surface embedded
in the flat Minkowski spacetime.
Our expression clearly satisfies this criterion, as can be seen by
the following observation\cite{abraham83}.
Let $\phi$ be a diffeomorphism from a given spacetime $M_{3+1}$ to itself,
and $\phi_{\ast}$ be the push-forward associated with $\phi$.
For any vector fields $X$, $Y$ defined on $M_{3+1}$, the inner product
and the Lie bracket are preserved by the mapping $\phi_{\ast}$\cite{abraham83},
\brr
& &   < \phi_{\ast}X, \ \phi_{\ast}Y>= \phi_{\ast}<X, \ Y>,  \label{sharkson}\\
& &  [\phi_{\ast}X, \ \phi_{\ast}Y ]_{\rm L}
=\phi_{\ast}[X, \ Y ]_{\rm L}.                    \label{deepsea}
\err
Suppose that $M_{3+1}$ is the flat Minkowski spacetime, and let
$n$ and $l$ be the null vector fields at each point of $N_{2}$
of the flat Minkowski spacetime.
Then these null vector fields remain null and the normalization condition is
preserved under the mapping $\phi_{\ast}$,
since we have
\brr
& & < \phi_{\ast}n, \ \phi_{\ast}n > =\phi_{\ast}<n, \ n >=0, \label{marine}\\
& & < \phi_{\ast}l, \ \phi_{\ast}l > =\phi_{\ast}<l, \ l >=0, \label{boy}\\
& & < \phi_{\ast}n, \ \phi_{\ast}l > =\phi_{\ast}<n, \ l >=-1.  \label{girl}
\err
The following identity
\be
[\phi_{\ast}n, \ \phi_{\ast}l ]_{\rm L}
=\phi_{\ast}[n, \ l ]_{\rm L}=0               \label{fishing}
\ee
is also true,
since two null vector fields at any points in the flat Minkowski spacetime
must commute.
Therefore, $[n, \ l ]_{\rm L}=0$ holds on any two-surface (and its deformations)
in the flat Minkowski spacetime. Thus, $L(u,v;\xi)$
is zero on any two-surface $N_{2}$ in the flat
Minkowski spacetime, modulo the subtraction term that can be
trivially put to zero.

Notice that $L(\xi)$ is linear in $\xi$. That is, for any
$\xi$ given by
\be
\xi = a\xi_{1} + b\xi_{2},
\ee
where $a,b$ are constants, we have
\be
L(\xi)=aL(\xi_1) + bL_(\xi_{2}),
\ee
which shows that the quasi-local angular momentum is additive.
Asymptotic properties of quasi-local angular momentum and its flux
will be studied in section \ref{k2}.

\end{subsection}

\begin{subsection}{Quasi-local rotational energy}{\label{geom4}}

Like quasi-local angular momentum, the value of any reasonable
candidate of quasi-local rotational energy must be zero for any two-surface
in the flat Minkowski spacetime.
The quasi-local rotational energy $W_{R}$
defined in (\ref{w1}) trivially satisfies this criterion. Since $W_{R}$
can be written as
\be
W_{R}
=-{1\over 16\pi}\int^{u}_{0}\!\! \! du \! \oint \! d^{2}y \,
A_{+}^{\ a}\partial_{+} \Big( {\rm e}^{ 2\sigma}\rho_{a b}
 [n, \ l ]_{\rm L}^{b} \Big),               \label{rotten}
\ee
it is zero when
\be
[n, \ l ]_{\rm L}^{a}=0,
\ee
which is true for any $n$ and $l$ of the flat Minkowski spacetime.

\end{subsection}

\end{section}

\begin{section}{Asymptotically flat limits}{\label{k2}}
\setcounter{equation}{0}

In this section I will show that the limits of
quasi-local balance equations in the asymptotically flat zones
are the Bondi energy-loss,
linear momentum-loss, and angular momentum-loss equations\cite{yoon01,yoon02}.
Moreover, the integral $W_{R}$ turns out to be proportional to the total angular
momentum squared in this limit, which is a strong indication that
it is to be regarded as a quasi-local generalization of the Carter's
constant\cite{carter68,hugh-pen72,carter79}
for a generic gravitational field.

The general form of asymptotically flat
metrics\cite{yoon01,wald84,penrose86,chrusciel02,prior77}
is given by
\begin{eqnarray}
ds^2 & & \longrightarrow
-2 du dv
-\Big( 1-{2m\over v} +\cdots \Big)\, du^2
+ \Big( {4 m a  {\rm sin}^{2}\vartheta \over v}
-{4 m a^{3}  {\rm sin}^{2}\vartheta   {\rm cos}^{2}\vartheta \over v^{3}}
 + \cdots \Big) du d \varphi           \nonumber\\
& &
+v^{2}
\Big( 1 + {a^2  {\rm cos}^{2}\vartheta \over v^{2}} + \cdots \Big)
d\vartheta^{2}
+v^{2}  {\rm sin}^{2}\vartheta
\Big( 1 + {a^2\over v^{2}} + \cdots \Big) d\varphi^{2}  \nonumber\\
& &
+  {\rm sin}^{2}\vartheta \Big(
{4 m a^3  \over v^{3}} +{8 m^{2} a^3  \over v^{4}}+ \cdots \Big)
dv d\varphi
-\Big( {a^2   {\rm sin}^{2}\vartheta \over v^{2}} + \cdots \Big)
dv^{2},                                          \label{sola}
\end{eqnarray}
as  $v \rightarrow \infty$. From this expansion,
asymptotic fall-off rates of the metric coefficients are found to be
\begin{eqnarray}
& &{\rm e}^{\sigma}=v^{2}   {\rm sin}\vartheta \Big\{
1 +O({1 \over v^{2}})  \Big\},           \label{fallsig}\\
& &\rho_{\vartheta \vartheta}
=({\rm sin}\vartheta)^{-1} \Big\{
1 + { C(u,\vartheta,\varphi)\over v}
+ O({1 \over v^{2}}) \Big\},       \label{fallrho1} \\
& &\rho_{\varphi \varphi}={\rm sin}\vartheta  \Big\{
    1 - { C(u,\vartheta,\varphi) \over v}
+ O({1 \over v^{2}}) \Big\},    \label{fallrho2} \\
& &\rho_{\vartheta \varphi}
=O({1\over v^{2}}),   \label{fallrho3}\\
& &2h=1-{2m \over v}
 + O({1 \over  v^{2}}),         \label{fallh}\\
& &A_{+}^{\ \varphi}={2ma \over v^{3}} + O({1 \over v^{4}}), \label{fallb}\\
& & A_{-}^{\ \varphi}= {2ma^{3} \over v^{5}}
+ O({1 \over v^{6}}),                      \label{fallc}\\
& & A_{\pm}^{\ \vartheta}=  O({1 \over v^{6}}),     \label{falla}
\end{eqnarray}
and their derivatives are given by
\begin{eqnarray}
& &
\partial_{+}\sigma = O({1\over v^{2}}), \label{peg}\\
& & \partial_{-}\sigma = {2\over v}+ O({1\over v^{2}}), \label{drill}\\
& & \partial_{+}\rho_{a b} = O({1\over v}),   \label{screw}\\
& & \partial_{-}\rho_{a b} = O({1\over v^{2}}), \label{stainless} \\
& & {\pounds}_{\xi} \rho_{a b} = O({1\over v}),  \label{brass}\\
& & \pi^{a b} = -{1\over 2}{\rm e}^{\sigma}\rho^{a c}\rho^{b d}
 (\partial_{+}\rho_{c d}) +  O(1),              \label{hammer}\\
& & \pi_{h}=-4v\,{\rm sin}\vartheta + O(1),  \label{mallet}\\
& & \pi_{\sigma}=-2v\, {\rm sin}\vartheta + O(1), \label{ruler}\\
& & \pi_{\varphi}= 6ma \,{\rm sin}^{3}\vartheta + O({1\over v}), \label{driver}\\
& & \pi_{\vartheta}= O({1\over v^{2}}).               \label{ducttape}
\end{eqnarray}
Therefore, $n$ and $l$ become, asymptotically,
\brr
& & n \longrightarrow  {\partial \over \partial u}
-  \Big({1\over 2}-{m\over v}\Big)
  {\partial \over \partial v},                            \label{cusp}\\
& & l \longrightarrow {\partial \over \partial v}.      \label{chisel}
\err
The natural reference spacetime at the asymptotic infinity
is the flat Minkowski spacetime,
\be
d\bar{s}^{2}=-2d\bar{u}d\bar{v} -d\bar{u}^{2} + r^{2}
(d \vartheta^{2} + {\rm sin}^{2}\vartheta d\varphi^{2}), \label{sailfish}
\ee
where $N_{2}=S_{2}$.
Thus the embedding degrees of freedom of $S_{2}$ into the flat Minkowski
spacetime are given by
\be
\bar{A}_{\pm}^{ \ \ a} = 0, \hspace{.2in}
2\bar{h} = 1,                \label{epoxy}
\ee
so that $\bar{n}$ and  $\bar{l}$ are given by
\brr
& &
\bar{n} = {\partial \over \partial \bar{u}}
- {1\over 2}  {\partial \over \partial \bar{v}} , \label{backn}\\
& &
\bar{l} = {\partial \over \partial \bar{v}}.  \label{backl}
\err

\begin{subsection}{The Bondi energy-loss relation}{\label{loss}}

In the asymptotic region $v\rightarrow \infty$ where $v={\rm constant}$
hypersurface is timelike, the r.h.s. of (\ref{enflux1})
represents the canonical energy-flux carried by gravitational radiation
crossing $N_{2}$.
Then the l.h.s. should be identified as
the instantaneous rate of change in the gravitational energy
of the region enclosed by $N_{2}$.
Energy-flux in general does not have a definite sign,
since it includes the energy-flux carried by the in-coming
as well as the out-going radiation across $N_{2}$.
In the asymptotically flat region, however,
energy-flux turns out to be negative-definite,
representing the physical situation that
there is no in-coming flux coming from the infinity\cite{BMS62}.

Let us show that the balance equation (\ref{enflux1}) indeed
reduces to the Bondi energy-loss formula at the null infinity.
Let $\bar{U}$ be given by (\ref{ubar}) and use $\bar{n}$ in
(\ref{backn}).
Then we find that
\be
\bar{U}={\bar{v}\over 2}.                            \label{flounder}
\ee
Let us suppose that the coordinates $\{u,v,y^{a}\}$
approach the coordinates of the Minkowski spacetime
$\{ \bar{u}, \bar{v}, \vartheta, \varphi \}$ as $v\rightarrow\infty$,
\be
u \longrightarrow \bar{u},   \hspace{.2in}
v \longrightarrow \bar{v},   \hspace{.2in}
y^{a} \longrightarrow \{ \vartheta, \varphi \}.         \label{arctic}
\ee
Then $\bar{U}$ trivially satisfies the condition (\ref{grouper}), since
\be
{\partial \over \partial u}\bar{U} \longrightarrow
{\partial \over \partial \bar{u}}\bar{U} = 0.       \label{octopus}
\ee
If we use the asymptotic formula
\be
n-\bar{n} \longrightarrow {m\over v}
{\partial \over \partial v},               \label{squid}
\ee
then the total energy is given by
\begin{eqnarray}
U_{\rm B}(u)
&:=&  \lim_{v \rightarrow \infty}U(u,v)
 =\lim_{v \rightarrow \infty}
{1 \over 8\pi} {\pounds}_{(n-\bar{n})}{\cal A} \nonumber\\
& =&  m(u),
\end{eqnarray}
which is just the Bondi mass at the null infinity.

Asymptotic limit of the balance equation (\ref{enflux1}) is found to be
\begin{eqnarray}
{d \over d u} U_{\rm B}(u)
& = &
-\lim_{v \rightarrow \infty}
{1\over 32\pi}  \oint_{S_{2}} \!\!\!
d \Omega \, \, v^{2}
\rho^{a c}\rho^{b d}(\partial_{+}\rho_{b c})
(\partial_{+}\rho_{a d})                 \nonumber\\
&=&
-\lim_{v \rightarrow \infty}
{1\over 32\pi}  \oint_{S_{2}} \!\!\!
d \Omega \, \, v^{2}(j^{\ a}_{ + \, b}j^{\ b}_{ + \, a})
\leq 0,                                 \label{lindoo}
\end{eqnarray}
where $j^{\ a}_{ + \, b}$ is the {\it shear current} defined as
\begin{equation}
j^{\ a}_{+ \, b}:=\rho^{a c}\partial_{+}\rho_{ b c} \hspace{.2in}
(j^{\ a}_{ + \, a}=0),            \label{tracu}
\end{equation}
which represents traceless shear degrees of freedom of gravitational radiation.
The relation (\ref{lindoo}) is just the Bondi energy-loss
formula with the correct normalization coefficient\cite{BMS62}.
Notice that the energy-flux is bilinear in $j^{\ a}_{+ \, b}$.
Equivalently, it can be written as
\begin{equation}
{d \over d u} U_{\rm B}(u)
=-{1\over 16\pi}  \oint_{S_{2}} \!\!\!
d \Omega \, \, (\partial_{+}C)^{2}
\leq 0,                            \label{lindo}
\end{equation}
if one uses the expansion of $\rho_{ab}$ given by
(\ref{fallrho1}), (\ref{fallrho2}), and (\ref{fallrho3}).

\end{subsection}

\begin{subsection}{The Bondi linear momentum and linear
momentum-flux}{\label{linear}}

The total linear momentum $P_{\rm B}(u)$ and its flux are trivially zero,
\brr
& &
P_{\rm B}(u) :=  \lim_{v\rightarrow \infty} P(u,v)
= -\lim_{v \rightarrow \infty}
{1 \over 8\pi} {\pounds}_{(l-\bar{l})}{\cal A} =0,   \label{bondim}\\
& &
{d \over d u} P_{\rm B}(u)=0,               \label{bondif}
\err
since we have
\be
l - \bar{l} \longrightarrow 0.               \label{cod}
\ee
That the net-flux of the total linear momentum is zero can be also seen
by evaluating each term of $H$ in (\ref{tilde}) in the asymptotic limit.
Let us notice that although the fourth, sixth, and seventh term in
(\ref{tilde}) are not zero individually, they add up to zero
asymptotically,
\brr
& & {1\over 2h}{\rm e}^{-\sigma}
\rho_{a b}\rho_{c d}\pi^{a c}\pi^{b d}
+{1\over 2h}\pi^{a b}(D_{+}\rho_{a b})
+{1\over 8h}{\rm e}^{\sigma}\rho^{a b} \rho^{c d}
(D_{+}\rho_{a c}) (D_{+}\rho_{b d})                 \nonumber\\
& = &
{h\over 2}{\rm e}^{\sigma}\rho^{a b} \rho^{c d}
(D_{-}\rho_{a c}) (D_{-}\rho_{b d})  \nonumber\\
& = &
O\Big(  {1\over v^{2} } \Big),       \label{fourseven}
\err
where we used the definition of $\pi^{a b}$ in (\ref{definition}).
The third and fifth terms become zero, respectively,
\brr
& & -{1\over 2}{\rm e}^{-2\sigma}\rho^{a b}\pi_{a}\pi_{b}
=O\Big(  {1\over v^{4}} \Big),              \label{worm}\\
& & {1\over 2}\pi_{h}(D_{+}\sigma)
=O\Big( {1\over v} \Big).                      \label{redworm}
\err
The remaining non-vanishing terms are given by
\brr
& &
\lim_{v\rightarrow \infty}
-{1 \over 16\pi} \oint_{S_{2}} \!\! \! d^{2}y \,  \Big(
{1\over 2}{\rm e}^{-\sigma}\pi_{h}\pi_{\sigma}
\Big)
=-1,                     \label{non2}\\
& &
\lim_{v\rightarrow \infty}
{1 \over 16\pi} \oint_{S_{2}} \!\!\!  d^{2}y \, \Big(
{1\over 4}h{\rm e}^{-\sigma}\pi_{h}^{2} \Big)
={1\over 2},              \label{non1}\\
& &
\lim_{v\rightarrow \infty}
{1 \over 16\pi} \oint_{S_{2}} \!\! \! d^{2}y \,
 {\rm e}^{\sigma}R_{2}
={1\over 4 } \chi,            \label{non3}
\err
where  $\chi=2$ for $S_{2}$. Since these terms add up to zero,
it follows that the net momentum-flux $H$ is zero at the null infinity.

\end{subsection}

\begin{subsection}{The Bondi angular momentum and angular
momentum-flux}{\label{angulars}}

Likewise, the total angular momentum $L_{\rm B}(u;\xi)$ is defined
as the asymptotic limit of quasi-local angular momentum $L(u,v;\xi)$,
\brr
L_{\rm B}(u;\xi)
&:=& \lim_{v\rightarrow \infty} L(u,v;\xi)     \nonumber\\
&=& -\lim_{v\rightarrow \infty}
{1 \over 16\pi} \oint \! d^{2}y \,
{\rm e}^{ \sigma} \Big( < \xi, \ [n, \ l ]_{\rm L}>
- < \xi, \ [\bar{n}, \ \bar{l} ]_{\rm L}>_{\rm ref} \Big). \label{deffa}
\err
Let $\xi$ be asymptotic to the azimuthal Killing vector field
\be
\xi =\xi^{a}\partial_{a}
\longrightarrow {\partial \over \partial \varphi}.
\ee
From the expansions (\ref{fallsig}), $\cdots$, (\ref{falla}), we find that
\brr
& & {\rm e}^{ \sigma}
\longrightarrow v^{2}{\rm sin}\vartheta,         \label{salmon}\\
& & \xi_{\varphi}
\longrightarrow v^{2}{\rm sin}^{2}\vartheta ,    \label{trolley}\\
& & \xi_{\vartheta}
\longrightarrow 0,                           \label{hook}
\err
where we used the definition of $\xi_{a}$ in (\ref{xidef}).
The Lie bracket $[n, \ l ]_{\rm L}$ is found to be
\brr
& &
[n, \ l ]_{\rm L}^{\varphi}\longrightarrow
-{6ma\over v^{4}}+ O\Big( {1\over v^{5}} \Big),        \label{line}\\
& &
[n, \ l ]_{\rm L}^{\vartheta}  \longrightarrow
O\Big( {1\over v^{6}} \Big).\label{northbrook}
\err
Since $[\bar{n}, \ \bar{l} ]_{\rm L}=0$, we find that
\brr
L_{\rm B}(u;\xi) &=&{1\over 16\pi}  \int_{0}^{2\pi} \! \! \! \! \!
d \varphi \! \int_{0}^{\pi} \! \! \! d \vartheta \,
(6ma) \, {\rm sin}^{3}\vartheta   \nonumber\\
&=&ma,
\err
which is just the total angular momentum of the Kerr spacetime.

The Bondi angular momentum-loss relation will be obtained by taking the limit
of the quasi-local balance equation (\ref{angflux}),
\be
{d L_{\rm B}\over du}=\lim_{v\rightarrow \infty}
{1 \over 16\pi}\oint_{S_{2}} \!\! \! d^{2}y \, \Big(
\pi^{a b}{\pounds}_{\xi} \rho_{a b}
+\pi_{\sigma}{\pounds}_{\xi} \sigma
-h {\pounds}_{\xi} \pi_{h}
-A_{+}^{\ a}{\pounds}_{\xi}\pi_{a} \Big).   \label{angfluxa}
\ee
Let us evaluate each term of this equation. The first term has
a finite limit, which is
\be
\oint_{S_{2}} \!\!\!
d^{2}y \,
\pi^{a b}{\pounds}_{\xi} \rho_{a b}
\longrightarrow
-{1 \over 2}\oint_{S_{2}} \!\!\!
d\Omega \, \, v^{2} \rho^{a c}\rho^{b d}
(\partial_{+}\rho_{c d}) ({\pounds}_{\varphi} \rho_{a b}), \label{first}
\ee
where we used the notation
\be
{\pounds}_{\varphi}:={\pounds}_{\partial / \partial \varphi}. \label{defvar}
\ee
Notice that
\be
\pi_{\sigma}{\pounds}_{\xi} \sigma
\longrightarrow
\Big\{ -2v \, {\rm sin}\vartheta + O(1) \Big\}
{\pounds}_{\varphi} \sigma.
\ee
But $\sigma$ becomes, asymptotically,
\be
\sigma  \longrightarrow
2{\rm ln}\, v + {\rm ln}\,|{\rm sin}\vartheta|
+ {\rm ln}\Big\{  1 + O\Big( {1\over v^{2}} \Big)   \Big\},
\ee
so that we have
\be
{\pounds}_{\varphi} \sigma =O\Big( {1\over v^{2}} \Big).
\ee
Thus the second term becomes zero,
\be
\oint_{S_{2}} \!\!\!
d^{2}y \,
\pi_{\sigma}{\pounds}_{\xi} \sigma
=O \Big( {1\over v} \Big) \longrightarrow 0.  \label{second}
\ee
Notice also that
\brr
h {\pounds}_{\xi} \pi_{h}
&=& {\pounds}_{\xi} (h \pi_{h}) -\pi_{h}  {\pounds}_{\xi} h  \nonumber\\
&\longrightarrow &
{\pounds}_{\xi} (   h \pi_{h} ) - 4 \,{\rm sin}\vartheta {\pounds}_{\xi} m
+ O\Big( {1\over v} \Big)           \nonumber\\
&=&
{\pounds}_{\xi} (   h \pi_{h} -  4 m \, {\rm sin}\vartheta )
+ O\Big( {1\over v} \Big).            \label{okterm}
\err
Thus the third term also becomes zero,
\be
\oint_{S_{2}} \!\!\!
d^{2}y \,
h {\pounds}_{\xi} \pi_{h}
=O \Big( {1\over v} \Big) \longrightarrow 0.  \label{third}
\ee
The fourth term dies off much faster, since
\be
\oint_{S_{2}} \!\!\!
d^{2}y \,
A_{+}^{\ a} {\pounds}_{\xi}\pi_{a}
=O({1\over v^{3}})
\longrightarrow 0.                 \label{fourth}
\ee
From (\ref{first}), (\ref{second}), (\ref{third}), and (\ref{fourth}),
we find that (\ref{angfluxa}) becomes
\be
{d L_{\rm B}\over du}
=-\lim_{v\rightarrow \infty}
{1\over 32\pi}\oint_{S_{2}} \!\!\!
d \Omega \,\, v^{2}
\rho^{a c}\rho^{b d}(\partial_{+}\rho_{c d})
({\pounds}_{\varphi} \rho_{a b}),          \label{aflux}
\ee
which is just the Bondi angular momentum-loss relation
with the correct normalization coefficient.
It is worth noting that (\ref{aflux}) is the
coordinate-dependent expression of the angular momentum-flux discussed
in \cite{ashtekar81,dray84,ashtekar91}.
If we use the asymptotic expansion of $\rho_{ab}$ given by
(\ref{fallrho1}), (\ref{fallrho2}), and (\ref{fallrho3}), then
this relation can be expressed as
\be
{d L_{\rm B}\over du}
=-{1\over 16\pi}\oint_{S_{2}} \!\!\! d \Omega \,\,
(\partial_{+}C)({\pounds}_{\varphi} C).    \label{totalflux}
\ee

\end{subsection}

\begin{subsection}{Gravitational Carter's constant }{\label{car}}

Let us find what $W_{R}$ becomes in this limit. The equation (\ref{w2})
becomes
\brr
{d \over d u}W_{R}
& \longrightarrow &
{3\over 4\pi v^{3}}  \int_{0}^{2\pi} \! \! \! \! \!
d \varphi \! \int_{0}^{\pi} \! \! \! d \vartheta \, \,
{\rm sin}^{3}\vartheta  \, (ma) \,
{d (ma) \over d u}   \nonumber\\
& = &
{1\over v^{3}} {d \over d u} (ma)^{2}.  \label{calca}
\err
If we choose the constant of $u$-integration as zero,
then we have
\be
\lim_{v\rightarrow \infty}v^{3}W_{R}=(ma)^{2},     \label{very}
\ee
which is just the total angular momentum squared
for the Kerr spacetime. Thus, $W_{R}$ may be regarded
as a quasi-local generalization of the Carter's
constant\cite{carter68,hugh-pen72,carter79,felice80},
and physically, could be interpreted as gravitational
contribution to the quasi-local rotational energy.

\end{subsection}

\end{section}

\vspace{.7cm}

\begin{section}{Relation between energy-loss and angular momentum-loss}
{\label{relation}}
\setcounter{equation}{0}

In general, if a given system undergoes an energy-losing process,
then it always accompanies angular momentum-loss,
unless the system remains spherically symmetric throughout the whole process.
Since we already found general expressions of quasi-local energy-flux and
angular momentum-flux, it is natural to ask what relation exists between them,
if there is any. The relation is that
the angular momentum-flux (\ref{angflux})
and energy-flux (\ref{enflux1}) transform into each other
\begin{equation}
{\partial \over \partial u}L(u,v;\xi) \longleftrightarrow
{\partial \over \partial u} U(u,v),           \label{cross}
\end{equation}
under the exchange of the derivatives
\be
{\pounds}_{\xi} \longleftrightarrow
{\partial \over \partial u}                \label{match}
\ee
in the flux integrals.

Let us examine the implication of this symmetry for spacetimes {\it close}
to a background spacetime that possesses two commuting
Killing vector fields. Let us choose the coordinates of the background spacetime
as $\{ \bar{u}, \bar{v}, \vartheta, \varphi\}$, and let
$\{ \partial / \partial \bar{u}, \partial / \partial \varphi \}$ be
two Killing vector fields of the background spacetime.
Let us also suppose that the coordinates $\{u,v,y^{a}\}$
approach the coordinates of the background spacetime
$\{ \bar{u}, \bar{v}, \vartheta, \varphi \}$ as $v\rightarrow\infty$,
\be
u \longrightarrow \bar{u},   \hspace{.2in}
v \longrightarrow \bar{v},   \hspace{.2in}
y^{a} \longrightarrow \{ \vartheta, \varphi \}.         \label{circa}
\ee

If we perturb this background spacetime by adding a small amount of
gravitational waves, then we may regard these waves as propagating
in the background spacetime, carrying a small amount of
energy and angular momentum. Let us write
\begin{math}
q^{I}=\{ h, \sigma,   A_{+} ^ { \ a}, \rho_{a b} \}
\end{math}
and
\begin{math}
\pi_{I}=\{ \pi_{h}, \pi_{\sigma},   \pi_{a}, \pi^{a b} \}
\end{math}
about an exact solution
$\{ \bar{q}^{I}, \bar{\pi}_{I}\}$
of the Einstein's equations,
\brr
& &
q^{I} = \bar{q}^{I}(\bar{v},\vartheta) + \delta q^{I},        \label{delq}\\
& &
\pi_{I} = \bar{\pi}_{I}(\bar{v},\vartheta) + \delta \pi_{I},
\err
where $\{ \delta q^{I}, \delta \pi_{I} \}$ represents gravitational waves
propagating on the background spacetime.
The dependence of a given mode on $\bar{u}$ and $\varphi$
is given by
\brr
& &
\delta q^{I}=Q^{I}(\bar{v},\vartheta)
\ {\rm e}^{i \omega \bar{u} + im_{z} \varphi} + {\rm c.c.}, \label{barq}\\
& &
\delta \pi_{I}=\Pi_{I}(\bar{v},\vartheta)
\ {\rm e}^{i \omega \bar{u} + im_{z} \varphi} + {\rm c.c.},  \label{barp}
\err
where  $\{ Q^{I}, \Pi_{I} \}$ are the amplitudes of the wave
that has the frequency $\omega$ and
the azimuthal quantum number $m_{z}$ ($m_{z}=0,\pm 1, \cdots$).
Now, if we use (\ref{circa}) and the fact that
$\{ \partial / \partial \bar{u}, \partial / \partial \varphi \} $
are the timelike and azimuthal
Killing vector fields of the background spacetime, respectively,
then from the symmetry (\ref{cross}) and (\ref{match})
we obtain the following relation\cite{bekenstein73}
\be
{d U \over d\bar{u}}={\omega \over m_{z}}{d L_{z} \over d\bar{u}},\label{aero}
\ee
which is a well-known relation between energy-loss and angular momentum-loss
for perturbations around the stationary and axi-symmetric spacetime.
Thus, the symmetry (\ref{cross}) and (\ref{match}) is the sought-for
relation between the energy-loss and angular momentum-loss for a generic
gravitational radiation that has no isometries.

\end{section}
\vspace{.7cm}

\begin{section}{Discussions} \label{discuss}
\setcounter{equation}{0}

The key ingredient of the (2,2) fibre bundle formalism discussed so far is
the observation that the out-going null vector field defines a natural time flow.
With the affine parameter of the out-going null geodesic as the time function
of the theory, the canonical variables were introduced and
the non-vanishing gravitational Hamiltonian was spelled out.
Then I obtained the Hamilton's equations of motion from the Hamiltonian
by the variational principle, which are the evolution equations along
the out-going null geodesics with respect to the affine parameter.
Thus, in this paper, I have shown that
the Einstein's equations split into twelve first-order Hamilton's
equations of motion and the four quasi-local balance
equations or constraint equations
that implement the Hamilton's evolution equations.

I also found coordinate-independent and geometric expressions of
quasi-local gravitational energy, linear momentum, and angular momentum
for any two-surface. The corresponding fluxes of
gravitational field were found to assume the canonical form of
energy-momentum flux,
\be
T_{0\alpha}\eta^{\alpha} \sim \sum_{I}\pi_{I}\pounds_{\eta}q^{I},
\ee
just as in standard field theories.
I have shown that the quasi-local balance equations correctly reproduce
the well-known Bondi relations at the null infinity
of asymptotically flat spacetimes. However, because of the breakdown
of the coordinate system due to potential occurrence of caustics
after a finite propagation along the out-going null geodesics,
there could be difficulties in extending the Hamilton's evolution
equations beyond the caustics.
In principle, however, it is still possible to approach the null infinity
by using a new coordinate system after the breakdown of the old one,
and perhaps one could use the quasi-local balance equations across
the caustics and search for ``weak'' solutions\cite{chorin-marsden}.
However, it must be mentioned that, the farther out one goes,
the less likely is the chance for caustics to occur
due to the weakness of gravity near the infinity.
If one is interested in the strong gravity region near black holes,
or black hole dynamics itself, then the caustics might cause serious problems
since they are much more likely to occur as we approach strong gravity region
along the in-going null geodesics.

Quasi-local angular momentum was defined in this paper for spacetimes
that have no isometries, and was found to be zero for any two-surface
in the flat Minkowski spacetime. It was found to have the additive property,
being a linear functional of a vector field $\xi$ that defines the
rotation at each point of the two-surface.
One might be interested in studying symmetry properties
of this quasi-local angular momentum, and look for some generalization of
the BMS symmetries at a finite region\cite{prior77}.

In addition, I obtained a quasi-local generalization of
the Carter's constant of gravitational field, and
interpreted it as gravitational contribution to the quasi-local
rotational energy.
The Carter's constant is known to exist when the system under study has
two commuting Killing vector fields, such as the Einstein-Maxwell system
and the Einstein's equations coupled to a scalar field. For a generic
gravitational field that has no isometries, no analog of
the Carter's constant is known. In this paper, I presented a candidate
for the generalized Carter's constant which becomes zero for any two-surface
in the flat Minkowski spacetime, and reduces to the total angular momentum
squared in the asymptotic region of the Kerr spacetime.
It is interesting to see how this quasi-local Carter's constant
generalizes when the electromagnetic and scalar fields
are present.

There could be a number of applications of the quasi-local balance
equations in astrophysics.
The most important and challenging problem seems to be
the calculation of back-reaction on the geometry of black holes
as a consequence of the emission or absorption of gravitational radiation.
One could also use the quasi-local balance
equations in searching for consistent boundary
conditions at a finite boundary in numerical relativity,
since  the boundary data at a finite boundary must
satisfy the quasi-local balance equations.
These problems are left for future works.

Another issue in this (2,2) formalism is the well-posedness
of the initial value problem.
When the initial 3 dimensional hypersurface
is chosen spacelike, there is
no problem in the well-posedness of the initial value problem
since the null direction can be viewed as a limit of timelike direction.
But there are several other choices of initial surfaces, such as the double null
initial surfaces and the initial/boundary value problems where
the boundary can be either timelike or null.
One of the difficulties associated with the characteristic
or initial/boundary value problem
is that one has to know the ``flows of information''
across the characteristic or the timelike boundary that belongs to the future.
In these hybrid formulations of
the Einstein's equations, not so many articles that aim to study the well-posedness
of the field equations appeared. However, a series of the recent papers
by Frittelli\cite{frittelli04} shows
that, for a certain choice of first-order variables for the characteristic problem
of the {\it linearized} Einstein's equations, the system can be cast into manifestly
well-posed form. For the non-linear characteristic problems, the notion of
well-posedness is still not available. It is interesting to examine whether
the first-order variables in this paper might have any relevance in establishing
the well-posedness of the non-linear characteristic initial value problem.

Finally, there are problems related to the
gauge invariance of this (2,2) fibre bundle formalism.
It is obvious that this formalism is tied to a particular gauge,
and the non-vanishing Hamiltonian is obtained as a consequence of
selecting a particular time function, namely,
choosing the affine parameter along the null direction as the time function.
But one should notice that, in the standard ADM formalism,
it is also possible to obtain another non-vanishing Hamiltonian
if one chooses a time function such as the Gauss normal
time coordinate\cite{kuchar92}.
Moreover, if one quantizes the theory in a particular gauge,
the resulting quantum theory will depend on that gauge, losing the
spacetime diffeomorphism invariance that one wishes to carry over
to the quantum regime. In view of the present situation that
there is not any single complete version of sensible quantum theory of gravity,
however, this gauge problem does not seem to be an urgent problem.
Clearly, quantizing the full Einstein's gravity is beyond the scope of
the present paper.

\end{section}

\bigskip\noindent
\centerline{\bf Acknowledgments}\\

It is a great pleasure to thank S. Hollands, R. Geroch
for interesting discussions at the last stage of preparing this paper,
and in particular, R. Wald for the hospitality extended to
the author during the sabbatical leave at the University of Chicago.
The author also thanks the referee for his criticism that helped to
improve the original manuscript on several delicate issues.

\newpage

\begin{appendix}

\renewcommand{\theequation}{\arabic{equation}}

\section*{Appendix: Hamilton's equations of motion}
\setcounter{equation}{0}

In this Appendix, I will show that, the Einstein's equations,
(\ref{cc}), (\ref{dd}), (\ref{fff}), and (\ref{ggg}), which are
second-order in $D_{-}$ derivatives, are the Hamilton's equations of motion,
\brr
& & D_{-}q^{I}=
  {\delta K \over \delta \pi_{I}}, \label{defmom}\\
& & D_{-}\pi_{I}=
  -{\delta K \over \delta q^{I}},   \label{defevol}
\err
if the boundary conditions
\be
\delta \sigma =\delta \rho_{a b} = 0       \label{bconn}
\ee
are assumed at the endpoints of $u$-integration in $K$,
where the Hamiltonian $K$ is given by
\be
K= \int \! \! du
\oint\!\! d^{2}y \, \Big\{
H+\lambda \, ({\rm det}\ \rho_{a b} - 1) \Big\},  \label{kagain}
\ee
and the Hamiltonian density $H$ is
\begin{eqnarray}
& & H  =  -{1\over 2}{\rm e}^{-\sigma}\pi_{\sigma}\pi_{h}
+ {1\over 4}h{\rm e}^{-\sigma}\pi_{h}^{2}
-{1\over 2}{\rm e}^{-2\sigma}\rho^{a b}\pi_{a}\pi_{b}
+{1\over 2h}{\rm e}^{-\sigma}
\rho_{a c}\rho_{b d}\pi^{a b}\pi^{c d}    \nonumber\\
& &
+{1\over 2}\pi_{h}(D_{+}\sigma)
+{1\over 2h}\pi^{a b}(D_{+}\rho_{a b})
+{1\over 8h}{\rm e}^{\sigma}\rho^{a b} \rho^{c d}
(D_{+}\rho_{a c}) (D_{+}\rho_{b d})
+{\rm e}^{\sigma}R_{2}.        \label{hagain}
\end{eqnarray}

\noindent
({\rm i}) \  {\it Variations with respect to $\pi{_h}$ and $h$}\\

It is trivial to see that the equation
\be
 D_{-}h= {\delta K\over \delta \pi_{h}}                \label{dmha}
\ee
is identical to the equation (\ref{dmh}), and the equation
\be
D_{-}\pi_{h}= - {\delta K \over \delta h}                \label{dmhb}
\ee
can be written as
\be
D_{-}\pi_{h}
 =   -{1\over 4}{\rm e}^{-\sigma}\pi_{h}^2
+{1\over 2h^2}{\rm e}^{-\sigma}\rho_{a b} \rho_{c d}
\pi^{a c}\pi^{b d}
+{1\over 2h^2}\pi^{a b}D_{+}\rho_{a b}
+{1\over 8h^2}{\rm e}^{\sigma} \rho^{a b}\rho^{c d}
(D_{+}\rho_{a c})(D_{+}\rho_{b d}).                 \label{dmpih}
\ee
Let us show that the equation (\ref{dmpih}) is just the equation (\ref{cc}),
using the equation (\ref{dmha}).
Notice that each term in (\ref{cc}) becomes
\brr
& {\rm (i)} &
2{\rm e}^{\sigma}D_{-}^{2}\sigma
   = -{1\over 2}{\rm e}^{-\sigma}
    \pi_{h}^2 -D_{-}\pi_{h},                   \label{h1}\\
& {\rm (ii)} & {\rm e}^{\sigma} (D_{-}\sigma)^{2}
={1\over 4}{\rm e}^{-\sigma}\pi_{h}^{2},         \label{hsec}\\
& {\rm (iii)} &
{1\over 2}{\rm e}^{\sigma}\rho^{a b}\rho^{c d} (D_{-}\rho_{a c})
(D_{-}\rho_{b d})
={1\over 2h^2}{\rm e}^{-\sigma}\rho_{a b} \rho_{c d}
\pi^{a c}\pi^{b d}
+{1\over 2h^2}\pi^{a b}D_{+}\rho_{a b}  \nonumber\\
& & \hspace{4.7cm}
+{1\over 8h^2}{\rm e}^{\sigma} \rho^{a b}\rho^{c d}
(D_{+}\rho_{a c})(D_{+}\rho_{b d}).          \label{hthird}
\err
From (\ref{h1}), (\ref{hsec}), and (\ref{hthird}), it follows that
the equation (\ref{cc}) is identical to the equation (\ref{dmpih}).
\\

\noindent
({\rm ii}) \  {\it Variations with respect to $\pi{_\sigma}$ and $\sigma$}\\

It is trivial to show that the equation
\be
D_{-}\sigma = {\delta K\over \delta \pi_{\sigma}}       \label{sigvar}
\ee
is identical to the equation (\ref{dmsig}). In the variation
\be
D_{-}\pi_{\sigma}= -{\delta K \over \delta \sigma}, \label{pisigvar}
\ee
the less trivial part is the following one,
\brr
\delta \int \! \! du \!
\oint \! d^{2}y \, ( \pi_{h}D_{+}\sigma )
& = &
\int \! \! du \!
\oint \! d^{2}y \, \pi_{h}D_{+}\delta \sigma  \no\\
& = &
-\int\! \! du  \!
\oint\! d^{2}y \, (D_{+} \pi_{h}) \delta \sigma
+ \int \! \! du  \!
\oint\! d^{2}y \, D_{+} (\pi_{h}\delta \sigma)  \no\\
& = &
-\int \! \! du  \!
\oint\! d^{2}y \, (D_{+} \pi_{h}) \delta \sigma
+ \int \! \! du  \! {d \over d u}
\Big\{ \oint \! d^{2}y \, \pi_{h}\delta \sigma  \Big\}.
\err
Therefore, if we assume the boundary condition
\be
\delta \sigma =0
\ee
at the endpoints of $u$-integration, then we have
\be
{1\over 2} {\delta \over \delta \sigma} \Big\{
\int
\! \! du \!
\oint \! d^{2}y \, \pi_{h}D_{+}\sigma   \Big\}
=-{1\over 2} D_{+}\pi_{h}.
\ee
The remaining variations are straightforward,
so that (\ref{pisigvar}) becomes
\brr
D_{-}\pi_{\sigma}
& = &
-{1\over 2}{\rm e}^{-\sigma}
\pi_{\sigma}\pi_{h}
+{1\over 4}h{\rm e}^{-\sigma}\pi_{h}^{2}
-{\rm e}^{-2\sigma} \rho^{a b}\pi_{a} \pi_{b}
+{1\over 2h}{\rm e}^{-\sigma}\rho_{a b} \rho_{c d}
\pi^{a c}\pi^{b d}        \nonumber\\
& &
+{1\over 2}D_{+}\pi_{h}
-{1\over 8h}{\rm e}^{\sigma} \rho^{a b}\rho^{c d}
(D_{+}\rho_{a c})(D_{+}\rho_{b d}).            \label{dmpisig}
\err
In order to show that the equation (\ref{dmpisig}) is the same as
the equation (\ref{fff}), we need to express
the derivatives of $D_{-}$ and $D_{-}^{2}$ in (\ref{fff}),
using the conjugate momenta.
Notice that the first term in (\ref{fff}) becomes
\brr
2{\rm e}^{\sigma}D_{-}^{2}h
& = &
{\rm e}^{\sigma} D_{-} \Big\{
-{\rm e}^{-\sigma}\pi_{\sigma}
+ D_{+}\sigma
+h{\rm e}^{-\sigma}\pi_{h} \Big\}  \no\\
& = &
-{\rm e}^{-\sigma}\pi_{h} \pi_{\sigma}
-D_{-}\pi_{\sigma}
+ {\rm e}^{\sigma}D_{-} D_{+}\sigma
+{1\over 2}\pi_{h}  D_{+}\sigma
+h{\rm e}^{-\sigma}\pi_{h}^{2}
+hD_{-}\pi_{h}.                     \label{dmh2}
\err
Since the third term in the r.h.s. of (\ref{dmh2}) can be written as
\brr
{\rm e}^{\sigma}D_{-}D_{+}\sigma
&=&
{\rm e}^{\sigma}D_{+}D_{-}\sigma
+\partial_{a}({\rm e}^{\sigma}F_{+-}^{\ \ a} ) \nonumber\\
&= &
-{1\over 2}D_{+}\pi_{h} + {1\over 2} \pi_{h}D_{+}\sigma
+\partial_{a} (  {\rm e}^{-\sigma} \rho^{a b}\pi_{b} ),       \label{que}
\err
(\ref{dmh2}) becomes
\brr
& {\rm (i)} &
2{\rm e}^{\sigma}D_{-}^{2}h
= -D_{-}\pi_{\sigma}
-{\rm e}^{-\sigma}\pi_{h} \pi_{\sigma}
-{1\over 2}D_{+}\pi_{h}
+\pi_{h}  D_{+}\sigma
+{3\over 4} h{\rm e}^{-\sigma}\pi_{h}^{2}
+{1\over 2h}{\rm e}^{-\sigma}\rho_{a b} \rho_{c d}
\pi^{a c}\pi^{b d}   \nonumber\\
& & \hspace{.74in}
+{1\over 2h}\pi^{a b}D_{+}\rho_{a b}
+{1\over 8h}{\rm e}^{\sigma} \rho^{a b}\rho^{c d}
(D_{+}\rho_{a c})(D_{+}\rho_{b d})
+\partial_{a}
(  {\rm e}^{-\sigma} \rho^{a b}\pi_{b} ),   \label{dmg1}
\err
where we used the equation of motion of $\pi_{h}$ given
by (\ref{dmpih}).
It is straightforward to express the remaining terms in (\ref{fff})
using the canonical variables.
They are given by
\brr
& {\rm (ii)} &
2{\rm e}^{\sigma} (D_{-}h) ( D_{-}\sigma )
={1\over 2}{\rm e}^{-\sigma}\pi_{h} \pi_{\sigma}
-{1\over 2}\pi_{h} D_{+}\sigma
-{1\over 2}h{\rm e}^{-\sigma}\pi_{h}^{2},    \label{dmg2}\\
& {\rm (iii)} &
{\rm e}^{\sigma}D_{+}D_{-}\sigma
= - {1\over 2} D_{+}\pi_{h}
+ {1\over 2} \pi_{h}D_{+}\sigma,          \label{dmg3}\\
& {\rm (iv)} &
{\rm e}^{\sigma}D_{-}D_{+}\sigma
=-{1\over 2}D_{+}\pi_{h} + {1\over 2} \pi_{h}D_{+}\sigma
+\partial_{a}
(  {\rm e}^{-\sigma} \rho^{a b}\pi_{b} ), \label{dmg4}\\
& {\rm (v)} & {\rm e}^{\sigma}(D_{+}\sigma)( D_{-}\sigma)
=-{1\over 2}\pi_{h} D_{+}\sigma,     \label{dmg5}\\
& {\rm (vi)} & {1\over 2}{\rm e}^{\sigma} \rho^{a b}\rho^{c d}
(D_{+}\rho_{a c})(D_{-}\rho_{b d})
={1\over 2h}\pi^{a b}D_{+}\rho_{a b}
+{1\over 4h}{\rm e}^{\sigma} \rho^{a b}\rho^{c d}
(D_{+}\rho_{a c})(D_{+}\rho_{b d}),        \label{dmg6}\\
& {\rm (vii)} &
{\rm e}^{2 \sigma}\rho_{a b}F_{+-}^{\ \ a}F_{+-}^{\ \ b}
={\rm e}^{-2 \sigma}\rho^{a b}\pi_{a}\pi_{b}.  \label{dmg7}
\err
If we plug (\ref{dmg1}), $\cdots$, (\ref{dmg7})
into (\ref{fff}), then the equation (\ref{fff}) becomes
\brr
& &
D_{-}\pi_{\sigma}
+{1\over 2}{\rm e}^{-\sigma}
\pi_{\sigma}\pi_{h}
-{1\over 4}h{\rm e}^{-\sigma}\pi_{h}^{2}
+{\rm e}^{-2\sigma} \rho^{a b}\pi_{a} \pi_{b}
-{1\over 2h}{\rm e}^{-\sigma}\rho_{a b} \rho_{c d}
\pi^{a c}\pi^{b d}    -{1\over 2}D_{+}\pi_{h}    \nonumber\\
& &
+{1\over 8h}{\rm e}^{\sigma} \rho^{a b}\rho^{c d}
(D_{+}\rho_{a c})(D_{+}\rho_{b d})
-2h{\rm e}^{\sigma}\Big\{ (D_{-}^{2}\sigma) +
{1\over 2}(D_{-}\sigma)^{2}
+ {1\over 4}\rho^{a b}\rho^{c d} (D_{-}\rho_{a c})
(D_{-}\rho_{b d}) \Big\}                \nonumber\\
& &
=0.                                      \label{dama}
\err
But the term in the bracket $\{\}$ in (\ref{dama}) is zero
if we use (\ref{cc}), and this shows that
the equations (\ref{fff}) and (\ref{dmpisig}) are identical.
\\

\noindent
({\rm iii}) \  {\it Variations with respect to $\pi_{a}$ and $A_{+}^{\ a}$}\\

The equation
\be
D_{-}A_{+}^{\ a}= {\delta K \over \delta \pi_{a}}
\ee
is
\be
D_{-}A_{+}^{\ a}=-{\rm e}^{-2 \sigma}\rho^{a b}\pi_{b}, \label{cova}
\ee
which is the defining equation (\ref{dmf}) of $\pi_{a}$,
since the covariant derivative of $A_{+}^{\ a}$ is given by
\brr
D_{-}A_{+}^{\ a}
& := & F_{-+}^{\ \ a}.      \label{cova1}
\err
In order to write down the equation
\be
D_{-}\pi_{a}=-{\delta K \over \delta A_{+}^{\ a}},  \label{covb}
\ee
one needs to do the following variations.
Notice that
\brr
\delta \int \! \! du \!
\oint \! d^{2}y \, ( \pi_{h}D_{+}\sigma )
& = & \int \! \! du \!
\oint \! d^{2}y \, \Big\{
\pi_{h} \delta \Big( \partial_{+}\sigma
- A_{+}^{\ a}\partial_{a}\sigma
-\partial_{a}A_{+}^{\ a} \Big) \Big\}  \no\\
& = &
\int \! \! du \!
\oint \! d^{2}y \, \Big\{
\Big(
-\pi_{h} \partial_{a}\sigma
+ \partial_{a}\pi_{h}  \Big) \delta A_{+}^{\ a}
-\partial_{a} \Big( \pi_{h}\delta A_{+}^{\ a}  \Big)
\Big\}.
\err
Thus we have
\be
 { \delta \over \delta A_{+}^{\ a } }
\int
\! \! du \!
\oint \! d^{2}y \,  {1\over 2} (\pi_{h}D_{+}\sigma)
=-{1\over 2} \pi_{h} \partial_{a}\sigma
+ {1\over 2} \partial_{a}\pi_{h}.   \label{vara1}
\ee
Let us also notice that, for an arbitrary field
$\zeta^{a b}$, the following is true,
\brr
& &
\int \! \! du \!
\oint \! d^{2}y \,
\zeta^{a b} \delta (D_{+} \rho_{a b})        \no\\
= & &
\int \! \! du \!
\oint \! d^{2}y \,
\zeta^{a b}
\Big\{
 - (\delta A_{+}^{\ c})\partial_{c}\rho_{a b}
-(\partial_{a}\delta A_{+}^{\ c})\rho_{c b}
-(\partial_{b}\delta A_{+}^{\ c})\rho_{a c}
+(\partial_{c}\delta A_{+}^{\ c}) \rho_{a b} \Big\} \no\\
= & &
\int \! \! du \!
\oint \! d^{2}y \, \Big[
\Big\{
-\zeta^{a b}\partial_{c}\rho_{a b}
+\partial_{a} ( \zeta^{a b}\rho_{c b})
+\partial_{b} ( \zeta^{a b}\rho_{a c})
-\partial_{c} ( \zeta^{a b}\rho_{a b}) \Big\}
\delta A_{+}^{\ c}           \no\\
& &
-\partial_{a} \Big( \zeta^{a b}\rho_{c b}\delta A_{+}^{\ c} \Big)
-\partial_{b} \Big( \zeta^{a b}\rho_{a c}\delta A_{+}^{\ c} \Big)
+\partial_{c} \Big( \zeta^{a b}\rho_{a b}\delta A_{+}^{\ c} \Big)
\Big].                    \label{multi}
\err
Let us consider the following two cases, (a) and (b).\\
(a) \ \ If we choose
\be
\zeta^{a b}:={1\over 2 h}\pi^{a b},
\ee
then from (\ref{multi}) we obtain
\be
{ \delta \over \delta A_{+}^{\ a } }
\int
\! \! du \!
\oint \! d^{2}y \,
{1\over 2 h}\pi^{b c}( D_{+}\rho_{b c} )
=-{1\over 2 h}\pi^{b c}  \partial_{a} \rho_{b c}
+ \partial_{b} \Big(
{1\over h}\pi^{b c}\rho_{c a} \Big),       \label{varaf1}
\ee
where we used the identity,
\be
\rho_{a b}\pi^{ab}=0.               \label{less}
\ee
(b) \ \ If we choose
\be
\zeta^{a b}:={1\over 4 h}{\rm e}^{\sigma}\rho^{a e}\rho^{b f}
(D_{+}\rho_{e f}),
\ee
which now depends on $A_{+}^{\ a }$, then (\ref{multi}) becomes
\brr
& &
{ \delta \over \delta A_{+}^{\ a } }
\int
\! \! du \!
\oint \! d^{2}y \,
{1\over 8 h}{\rm e}^{\sigma}\rho^{b d}\rho^{c e}
(D_{+}\rho_{b c})( D_{+}\rho_{d e})              \nonumber\\
& = &
 -{1\over 4 h}{\rm e}^{\sigma}\rho^{b d}\rho^{c e}
(D_{+}\rho_{d e}) (\partial_{a} \rho_{b c} )
+\partial_{b}
\Big( {1\over 2 h}{\rm e}^{\sigma}\rho^{b c}D_{+}\rho_{c a}
\Big),                  \label{varaf2}
\err
where we used the identity
\be
\rho^{a b}D_{+}\rho_{a b}=0.
\ee
From the equations (\ref{vara1}), (\ref{varaf1}), and
(\ref{varaf2}), we find that the equation (\ref{covb})
becomes
\brr
D_{-}\pi_{a}
& = & {1\over 2}\pi_{h}\partial_{a}\sigma
-{1\over 2}\partial_{a}\pi_{h}
+ {1\over 2h}
\Big\{ \pi^{b c} +{1\over 2}{\rm e}^{\sigma}
\rho^{b d}\rho^{c e}(D_{+}\rho_{d e}) \Big\}
(\partial_{a}\rho_{bc})                 \nonumber\\
& &
-\partial_{b}\Big\{
 {1\over h}\pi^{b c} \rho_{c a}
+ {1 \over 2 h}
{\rm e}^{\sigma}\rho^{b c}(D_{+}\rho_{c a}) \Big\}.  \label{gooda}
\err
Using the definitions of the momenta
(\ref{dmh}), $\cdots$, (\ref{dmrho}), one can
easily show that the equation (\ref{dd}) is identical to (\ref{gooda}).
\\

\noindent
({\rm iv}) \   {\it Variations with respect to $\pi^{ab}$ and $\rho_{ab}$}\\

It is trivial to see that the equation
\be
D_{-}\rho_{a b} = {\delta K \over \delta \pi^{a b}}   \label{covp1}
\ee
is just the equation (\ref{dmrho}). Let us show that the equation
\be
D_{-}\pi^{a b}=-{\delta K \over \delta \rho_{a b} }  \label{covp2}
\ee
is identical to the equation (\ref{ggg}).
If we vary terms in $K$ which do not contain $D_{+}\rho_{ab}$, then we have
\be
{ \delta \over \delta \rho_{a b} }
\int \! \! du \!
\oint \! d^{2}y \,  \Big(
-{1\over 2}{\rm e}^{-2\sigma}\rho^{c d}\pi_{c}\pi_{d}
+{1\over 2h}{\rm e}^{-\sigma}
\rho_{c e}\rho_{d f}\pi^{c e}\pi^{d f} \Big)
= {1\over 2}{\rm e}^{-2\sigma}\rho^{a c}\rho^{b d}\pi_{c}\pi_{d}
+{1\over h}{\rm e}^{-\sigma}
\rho_{c d}\pi^{a c}\pi^{b d}.            \label{rhovar1}
\ee
Varying terms linear in $D_{+}\rho_{ab}$, we find that
\be
\delta \int \! \! du \!
\oint \! d^{2}y \, \Big( {1\over 2h}\pi^{a b}D_{+}\rho_{a b}\Big)
= -\int \! \! du \!
\oint \! d^{2}y \,
D_{+}\Big(  {1\over 2h}\pi^{a b} \Big)\delta \rho_{a b}
+\int \! \! du \! {d \over d u} \Big\{
\oint \! d^{2}y \,
\Big( {1\over 2h}\pi^{a b}\delta \rho_{a b}  \Big)  \Big\}.   \label{ara}
\ee
If we assume the boundary condition
\be
\delta \rho_{a b}=0,                      \label{bcrho}
\ee
then we have
\be
{\delta \over \delta \rho_{a b}}
\int \! \! du \!
\oint \! d^{2}y \, \Big(
{1\over 2h}\pi^{c d}D_{+}\rho_{c d}\Big)
=-D_{+}\Big(  {1\over 2h}\pi^{a b} \Big).    \label{rhovar2}
\ee

Now let us define
\be
S^{a}_{\ b}:= \rho^{a c}D_{+}\rho_{c b}.
\ee
Then we have
\be
{1\over 8h}{\rm e}^{\sigma}\rho^{a b} \rho^{c d}
(D_{+}\rho_{a c}) (D_{+}\rho_{b d})
={1\over 8h}{\rm e}^{\sigma}
S^{b}_{\ a}\, S^{a}_{\ b},
\ee
and the variation of $S^{a}_{\ b}$ is given by
\be
\delta S^{a}_{\ b}
= -\rho^{a c}S^{d}_{\ b}\delta \rho_{cd}
+\rho^{a c}D_{+}\delta \rho_{c b}.
\ee
Therefore, we have
\brr
& & \delta \int \! \! du \!
\oint \! d^{2}y \, \Big\{
{1\over 8h}{\rm e}^{\sigma}\rho^{a b} \rho^{c d}
(D_{+}\rho_{a c}) (D_{+}\rho_{b d}) \Big\}       \nonumber\\
& = & \int \! \! du \!
\oint \! d^{2}y \, \Big(
{1\over 4h}{\rm e}^{\sigma}
S^{b}_{\ a}\, \delta  S^{a}_{\ b}\Big) \hspace{1in} \no\\
& = &
\int \! \! du \!
\oint \! d^{2}y \, \Big\{
-{1\over 4h}{\rm e}^{\sigma} \rho^{a c}
  S^{d}_{\ c} \, S^{b}_{\ d} \, \delta \rho_{a b}
+{1\over 4h}{\rm e}^{\sigma} \rho^{a c}
 S^{b}_{\ c} D_{+}\delta \rho_{a b}  \Big\}  \no\\
& = &
\int \! \! du \!
\oint \! d^{2}y \, \Big\{
-{1\over 4h}{\rm e}^{\sigma} \rho^{a c}
  S^{d}_{\ c}  \ S^{b}_{\ d}
-D_{+}\Big( {1\over 4h}{\rm e}^{\sigma} \rho^{a c}
 S^{b}_{\ c} \Big) \Big\} \delta \rho_{a b}.
\err
Therefore, we find that
\brr
{ \delta \over \delta \rho_{a b} }
\int \! \! du \!
\oint \! d^{2}y \,  \Big\{
{1\over 8h}{\rm e}^{\sigma}\rho^{c e} \rho^{d f}
(D_{+}\rho_{c d}) (D_{+}\rho_{e f}) \Big\}
& = &
-{1\over 4h}{\rm e}^{\sigma} \rho^{a c}\rho^{b d}\rho^{e f}
(D_{+}\rho_{c e})(D_{+}\rho_{d f})         \nonumber\\
& &
-D_{+}\Big( {1\over 4h}{\rm e}^{\sigma} \rho^{a c}
\rho^{b d} D_{+}\rho_{c d}  \Big).   \label{rhovar3}
\err
Finally, we have to vary the Lagrange multiplier term in (\ref{fundam}).
It is given by
\be
{ \delta \over \delta \rho_{a b} }
\int \! \! du \!
\oint \! d^{2}y \,
\lambda \, ({\rm det}\ \rho_{c d} - 1)
= \lambda \, \rho^{a b}.                   \label{mulcon}
\ee
The scalar curvature term ${\rm e}^{\sigma}R_{2}$ is a topological density that
does not contribute to the metric variation. From (\ref{rhovar1}),
(\ref{rhovar2}), (\ref{rhovar3}), and (\ref{mulcon}), we have
\brr
& &
D_{-}\pi^{a b}
+ {1\over 2}{\rm e}^{-2\sigma}\rho^{a c}\rho^{b d}\pi_{c}\pi_{d}
+{1\over h}{\rm e}^{-\sigma}
\rho_{c d}\pi^{a c}\pi^{b d}
-D_{+}\Big\{
{1\over 2h}\pi^{a b}
+ {1\over 4h}{\rm e}^{\sigma} \rho^{a c}
\rho^{b d} (D_{+}\rho_{c d})  \Big\}        \no\\
& &
-{1\over 4h}{\rm e}^{\sigma} \rho^{a c}\rho^{b d}\rho^{e f}
(D_{+}\rho_{c e})(D_{+}\rho_{d f})
+ \lambda \, \rho^{a b}  = 0.  \label{hasto}
\err
The Lagrange multiplier $\lambda$ is determined by taking the trace of
(\ref{hasto}).
Notice that for any traceless field  $\chi^{a b}$ such that
\be
\rho_{ab}\chi^{a b}=0,
\ee
we have
\be
\rho_{ab}D_{\pm}\chi^{a b}=- \chi^{a b}D_{\pm}\rho_{ab}.
\ee
Thus, for $\chi^{a b}$ defined as
\be
\chi^{a b}:={1\over 2h}\pi^{a b}
+ {1\over 4h}{\rm e}^{\sigma} \rho^{a c} \rho^{b d} D_{+}\rho_{c d},
\ee
one has
\be
-\rho_{ab}
D_{+}\Big( {1\over 2h}\pi^{a b}
+ {1\over 4h}{\rm e}^{\sigma} \rho^{a c} \rho^{b d} D_{+}\rho_{c d}
\Big)
 =  {1\over 2h}\pi^{a b}D_{+}\rho_{a b}
+{1\over 4h}{\rm e}^{\sigma} \rho^{a c} \rho^{b d} (D_{+}\rho_{ab})
(D_{+}\rho_{c d}).                \label{tata}
\ee
Therefore, the trace of the equation (\ref{hasto}) becomes
\brr
0
&= &
2\lambda -\pi^{ab}D_{-}\rho_{ab}
+{1\over 2}{\rm e}^{-2\sigma}\rho^{a b}\pi_{a}\pi_{b}
+{1\over h}{\rm e}^{-\sigma}
\rho_{a b}\rho_{c d}\pi^{a c}\pi^{b d}
+{1\over 2h}\pi^{a b}D_{+}\rho_{a b} \nonumber\\
&=&
2\lambda +{1\over 2}{\rm e}^{-2\sigma}\rho^{a b}\pi_{a}\pi_{b}
+\pi^{ab} \Big(
-D_{-}\rho_{ab}
+{1\over h}{\rm e}^{-\sigma}
\rho_{a c}\rho_{b d}\pi^{c d}
+{1\over 2h}D_{+}\rho_{a b}   \Big) \nonumber\\
& = &
2\lambda +{1\over 2}{\rm e}^{-2\sigma}\rho^{a b}\pi_{a}\pi_{b}, \label{how}
\err
where we used the equation (\ref{covp1}). Thus $\lambda$ is given by
\be
\lambda = -{1\over 4}{\rm e}^{-2\sigma}
\rho^{c d}\pi_{c}\pi_{d}.              \label{detmind}
\ee
Thus we find that
the equation (\ref{hasto}) finally becomes
\brr
& & D_{-}\pi^{a b}
 =  -{1\over 2}{\rm e}^{-2\sigma}
\rho^{a c}\rho^{b d}\pi_{c} \pi_{d}
+{1\over 4}{\rm e}^{-2\sigma}
\rho^{a b}\rho^{c d}\pi_{c} \pi_{d}
-{1\over h}{\rm e}^{-\sigma}\rho_{c d}
\pi^{a c}\pi^{b d}              \nonumber\\
& & \hspace{0.5cm}
+D_{+}\Big\{ {1\over 2h}\pi^{a b}
+ {1\over 4h}{\rm e}^{\sigma}\rho^{a c}\rho^{b d}
(D_{+}\rho_{c d} ) \Big\}
+{1\over 4h}{\rm e}^{\sigma}\rho^{a c}\rho^{b d}\rho^{e f}
(D_{+}\rho_{c e})(D_{+}\rho_{d f}).        \label{done}
\err
One can show that this equation is the same as the equation (\ref{ggg}).
To show this, let us multiply the equation (\ref{done})
by $\rho_{a m}\rho_{b n}$, and use the definitions of the conjugate momenta
(\ref{pisigma}), $\cdots$, (\ref{definition}).
Then each term in (\ref{done}) becomes as follows,
\brr
({\rm i})&  &
\rho_{a m}\rho_{b n} (D_{-}\pi^{a b})           \nonumber\\
& = & \rho_{a m}\rho_{b n}D_{-}\Big\{
-{1\over 2}{\rm e}^{\sigma}\rho^{a c}\rho^{b d}
( D_{+}\rho_{c d} )
+h{\rm e}^{\sigma}\rho^{a c}\rho^{b d}
( D_{-}\rho_{c d} )    \Big\}                  \nonumber\\
& = & -{1\over 2}{\rm e}^{\sigma} (D_{-}\sigma) ( D_{+}\rho_{mn} )
+{1\over 2}{\rm e}^{\sigma}
\rho^{c d}( D_{-}\rho_{m c})( D_{+}\rho_{n d} )
+{1\over 2}{\rm e}^{\sigma} \rho^{c d}( D_{-}\rho_{n c})( D_{+}\rho_{m d} ) \no\\
& &
-{1\over 2}{\rm e}^{\sigma} ( D_{-}D_{+}\rho_{mn} )
+ {\rm e}^{\sigma} ( D_{-}h) (D_{-}\rho_{mn} )
+ h {\rm e}^{\sigma} (D_{-}\sigma) ( D_{-}\rho_{mn} ) \no\\
& &
-2h{\rm e}^{\sigma} \rho^{c d}( D_{-}\rho_{m c})( D_{-}\rho_{n d} )
+h {\rm e}^{\sigma} (D_{-}^{2}\rho_{mn} ),             \label{ash}\\
({\rm ii})& &
{1\over 2}{\rm e}^{-2\sigma} \pi_{m} \pi_{n}
-{1\over 4}{\rm e}^{-2\sigma}
\rho_{m n}\rho^{c d}\pi_{c} \pi_{d}
= {1\over 2}{\rm e}^{2 \sigma}\rho_{m c}\rho_{n d}
F_{+-}^{\ \ c}F_{+-}^{\ \ d}
-{1\over 4}{\rm e}^{2 \sigma}\rho_{m n}\rho_{c d}
F_{+-}^{\ \ c}F_{+-}^{\ \ d},                    \label{maple}\\
({\rm iii})& &
{1\over h}{\rm e}^{-\sigma}\rho_{c d}\pi^{a c}\pi^{b d}
\rho_{am}\rho_{bn}                \nonumber\\
& = &
{1\over 4h}{\rm e}^{\sigma} \rho^{c d}( D_{+}\rho_{m c})( D_{+}\rho_{n d} )
-{1\over 2}{\rm e}^{\sigma} \rho^{c d}
( D_{+}\rho_{m c})( D_{-}\rho_{n d} )
-{1\over 2}{\rm e}^{\sigma} \rho^{c d}( D_{+}\rho_{n c})
( D_{-}\rho_{m d} )                  \nonumber\\
& &
+h{\rm e}^{\sigma} \rho^{c d}( D_{-}\rho_{m c})( D_{-}\rho_{n d} ), \label{butter}\\
({\rm iv})&  &
-\rho_{a m}\rho_{b n} D_{+}\Big\{
{1\over 2h}\pi^{a b}
+ {1\over 4h}{\rm e}^{\sigma}
\rho^{a c}\rho^{b d}(D_{+}\rho_{c d}) \Big\}     \nonumber\\
& = & -{1\over 2}\rho_{a m}\rho_{b n} D_{+} \Big\{
{\rm e}^{\sigma}\rho^{a c}\rho^{b d}(D_{-}\rho_{c d}) \Big\} \no\\
& = &  - {1\over 2}{\rm e}^{\sigma} (D_{+}\sigma) ( D_{-}\rho_{mn} )
+{1\over 2} {\rm e}^{\sigma}\rho^{c d}( D_{+}\rho_{m c})( D_{-}\rho_{n d} )
+{1\over 2} {\rm e}^{\sigma}\rho^{c d}( D_{+}\rho_{n c})
( D_{-}\rho_{m d} )                   \nonumber\\
& &
-{1\over 2} {\rm e}^{\sigma}( D_{+}D_{-}\rho_{m n}), \label{nut}\\
({\rm v}) & &
-{1\over 4h}{\rm e}^{\sigma} \rho^{a c}\rho^{b d}\rho^{e f}
( D_{+}\rho_{c e})( D_{+}\rho_{d f} )\rho_{a m}\rho_{b n}
=-{1\over 4h}{\rm e}^{\sigma} \rho^{c d}
( D_{+}\rho_{m c})( D_{+}\rho_{n d} ).          \label{rhoab}
\err
After a little algebra, we find that the equation (\ref{done}) is
identical to the equation (\ref{ggg}).
Thus, assuming the boundary conditions (\ref{bconn}),
I have shown that the twelve Hamilton's equations of motion,
(\ref{defmom}) and (\ref{defevol}), are just the first-order form
of the six Einstein's equations (\ref{cc}), (\ref{dd}), (\ref{fff}),
and (\ref{ggg}). Therefore, the Hamilton's equations of motion,
(\ref{defmom}) and (\ref{defevol}), together with the four divergence-type
equations (\ref{qenergy}), (\ref{qmomentum}), and (\ref{qangular}),
are completely equivalent to the full Einstein's equations
(\ref{aa}), $\cdots$, (\ref{ggg}).

\end{appendix}

\nopagebreak

\end{document}